%% file: bitpath.tex
\documentclass{llncs}
\usepackage{makeidx}  
\usepackage{graphicx}
\usepackage{epsfig}
\usepackage[noend]{algorithmic}
\usepackage{algorithm}
\usepackage{url}
\usepackage{subfigure}
\usepackage{epsfig}
\begin{document}
\frontmatter          
\pagestyle{headings}  
\mainmatter              
\title{BitPath -- Label Order Constrained Reachability Queries over Large Graphs}
\author{Medha Atre\inst{1}, Vineet Chaoji\inst{2}, Mohammed J. Zaki\inst{3}}
\institute{University of Pennsylvania, Philadelphia PA, USA\\
\email{atrem@cis.upenn.edu}
\and
Yahoo! Labs, Bangalore, India\\
\email{chaojv@yahoo-inc.com}
\and
Rensselaer Polytechnic Institute, Troy NY, USA\\
\email{zaki@cs.rpi.edu}
}

\maketitle              
\begin{abstract}
In this paper we focus on the following constrained reachability problem over
edge-labeled graphs like RDF --
\textit{given source node x, destination node y, and a sequence of
edge labels (a, b, c, d), is there a
path between the two nodes such that the edge labels on the path satisfy a regular
expression ``*a.*b.*c.*d.*''}. A ``*'' before ``\textit{a}'' allows any other edge label
to appear on the path before edge \textit{``a}''. ``\textit{a.*}'' forces at least one
edge with label ``\textit{a}''. ``.*'' after ``\textit{a}'' allows zero or more edge labels after
``\textit{a}'' and before ``\textit{b}''.
Our query processing algorithm uses simple \textit{divide-and-conquer} and
\textit{greedy pruning} procedures to limit the search space. However, our graph
indexing technique -- based on \textit{compressed bit-vectors} -- allows
indexing large graphs which otherwise would have been infeasible.
We have evaluated
our approach on graphs with more than 22 million edges and 6 million nodes --
much larger compared to the datasets used in the contemporary work on path queries.

\end{abstract}

\input{intro}

\input{related}

\input{index}

\input{query_algo}

\input{eval}

\input{conclusion}

\bibliographystyle{abbrv}
\bibliography{bitpathbib}  

\begin{figure}
	\centering
		\includegraphics[width=5in,height=2in]{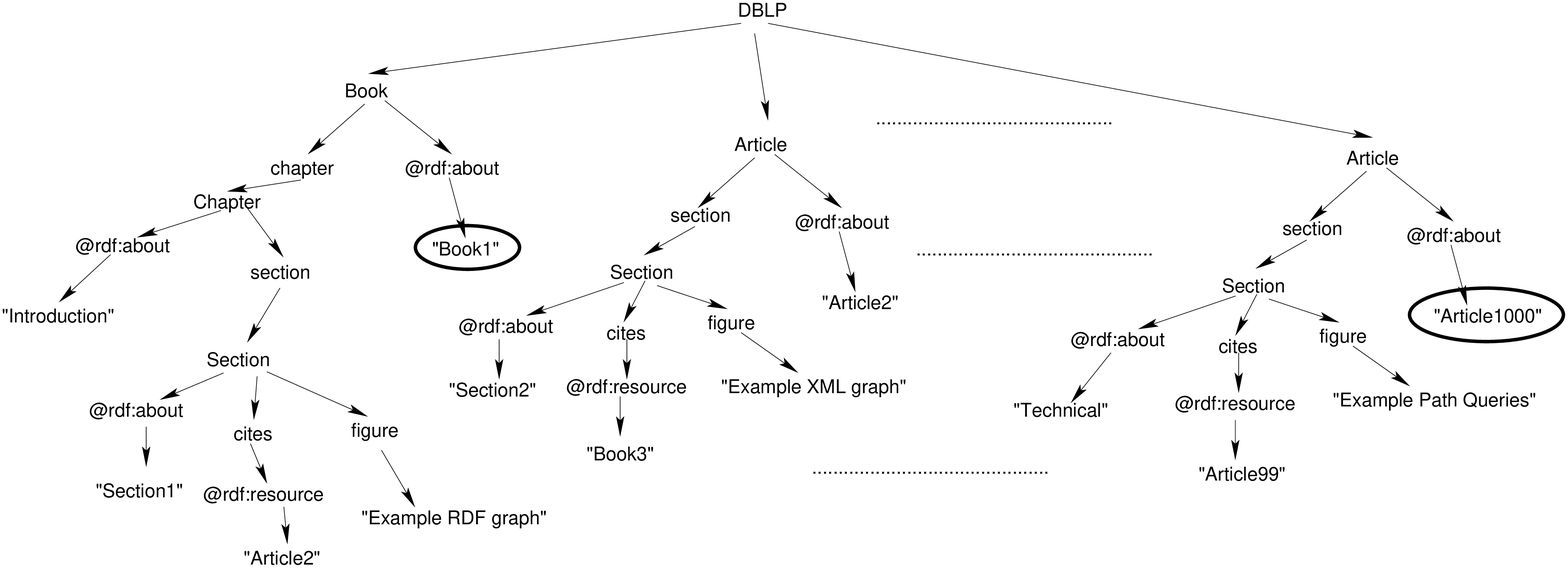}
	\caption{Example XML representation of RDF graph of DBLP data}
	\label{fig:xml}
\end{figure}

\appendix
\section{LOCR Query for XML representation of RDF} \label{apdx:rdfxml}
\begin{figure}[ht]
\begin{minipage}[b]{0.5\linewidth}
\centering
\includegraphics[scale=0.3]{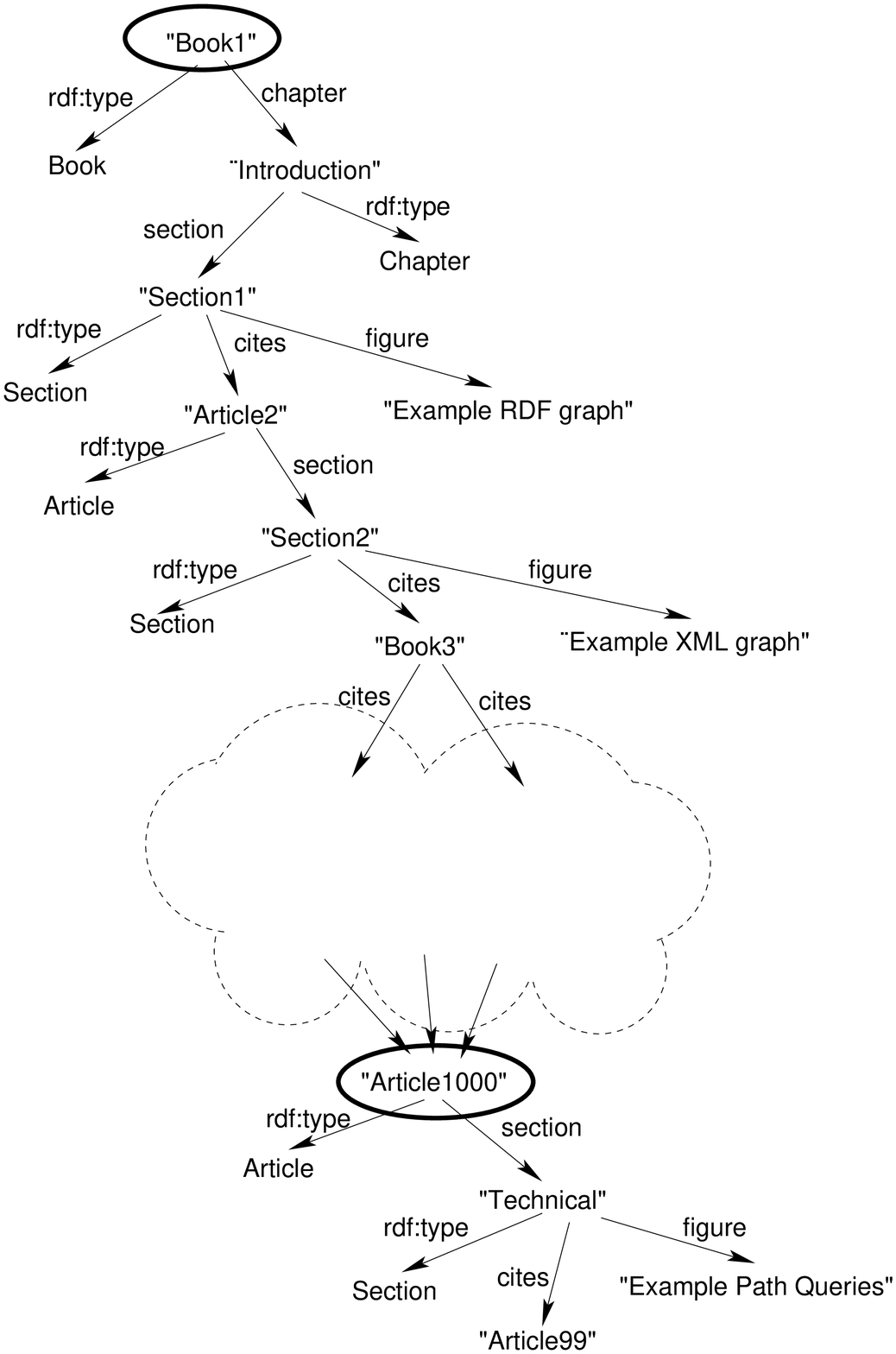}
\end{minipage}
\hspace{0.5cm}
\begin{minipage}[b]{0.5\linewidth}
\begin{scriptsize}
\begin{verbatim}
<DBLP>
<Book rdf:about="Book1">
 <chapter>
  <Chapter rdf:about="Introduction">
   <section>
    <Section rdf:about="Section1">
     <cites rdf:resource="Article2"/>
     <figure>"Example RDF graph"</figure>
    </Section>
  </section>
  </Chapter>
 </chapter>
</Book>
<Article rdf:about="Article2">
  <section>
   <Section rdf:about="Section1">
    <cites rdf:resource="Book3"/>
    <figure>"Example XML graph"</figure>
   </Section>
  </section>
</Article>
..............
<Article rdf:about="Article1000">
 <section>
  <Section rdf:about="Conclusion">
    <cites rdf:resource="Article99"/>
    <figure>"Example Path Queries"</figure>
  </Section>
 </section>
</Article>
</DBLP>
\end{verbatim}
\end{scriptsize}
\end{minipage}
\caption{Example RDF graph of DBLP data}\label{fig:rdfxml}
\end{figure}

Consider texual RDF/XML representation of a DBLP\footnote{\scriptsize{\url{http://www.w3.org/TR/rdf-syntax-grammar/}}} dataset
given in Figure \ref{fig:rdfxml}.
For simplicity we have three sample entries -- ``Book1'', ``Article2'', and ``Article1000'',
although the dataset can contain several such entries, denoted by ``......'' in the example.
The XML tree representation of this dataset with the three entries is shown in Figure \ref{fig:xml}.
Note that by default RDF to XML conversion tools do not add ID/IDREFs to cross reference same URIs
in the XML data.
The corresponding RDF graph is shown in Figure \ref{fig:rdfxml}.
The \textit{cloud} shown in the figure represents many more edges and nodes in between the citations.

Suppose the RDF graph has a transitive path with edges labeled ``cites'' from
element ``Book1'' to element ``Article1000''.
We are interested in the following LOCR query, (Book1, Article1000, (cites)). Looking at the XML
representation of the RDF graph in Figure \ref{fig:xml}, there is no such path between element ``Book1''
and ``Article1000'', although such a path may exist between corresponding nodes in the RDF graph.
This example shows that an LOCR query cannot always be translated into an equivalent path 
query over XML graph.

\section{Distribution of Queries per Query Length} \label{apdx:qcnt_q_size}
\begin{figure}[!ht]
 \centerline{
    \subfigure{
	\includegraphics[scale=0.35]{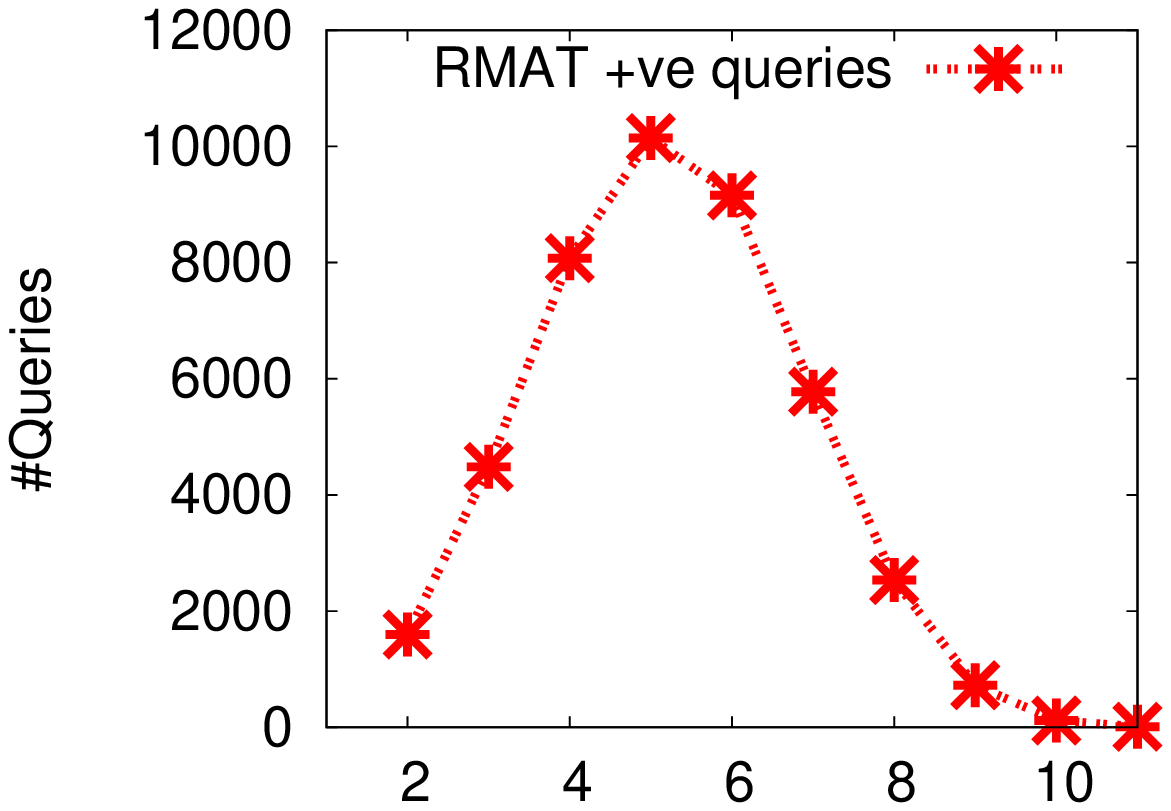}
    }
  \hspace{-0.2in}
    \subfigure{
	\includegraphics[scale=0.35]{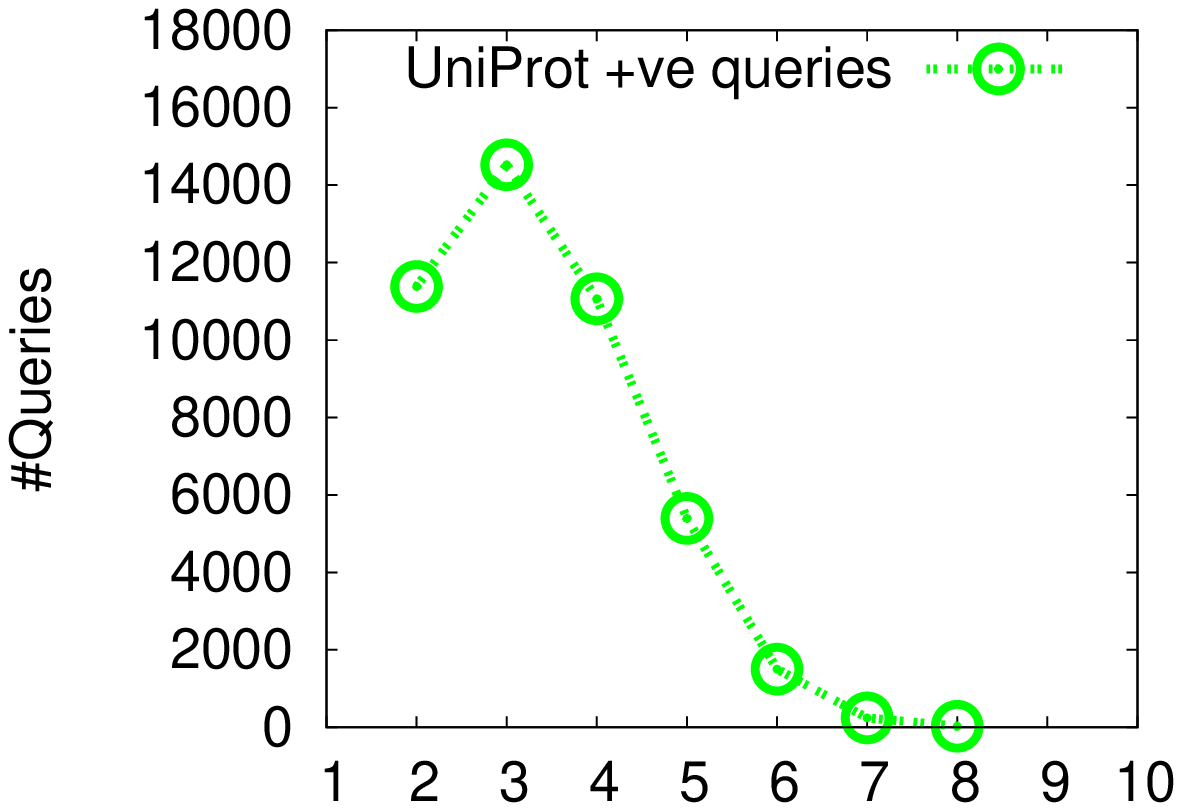}
    } \hspace{-0.2in}
   \subfigure{
	\includegraphics[scale=0.35]{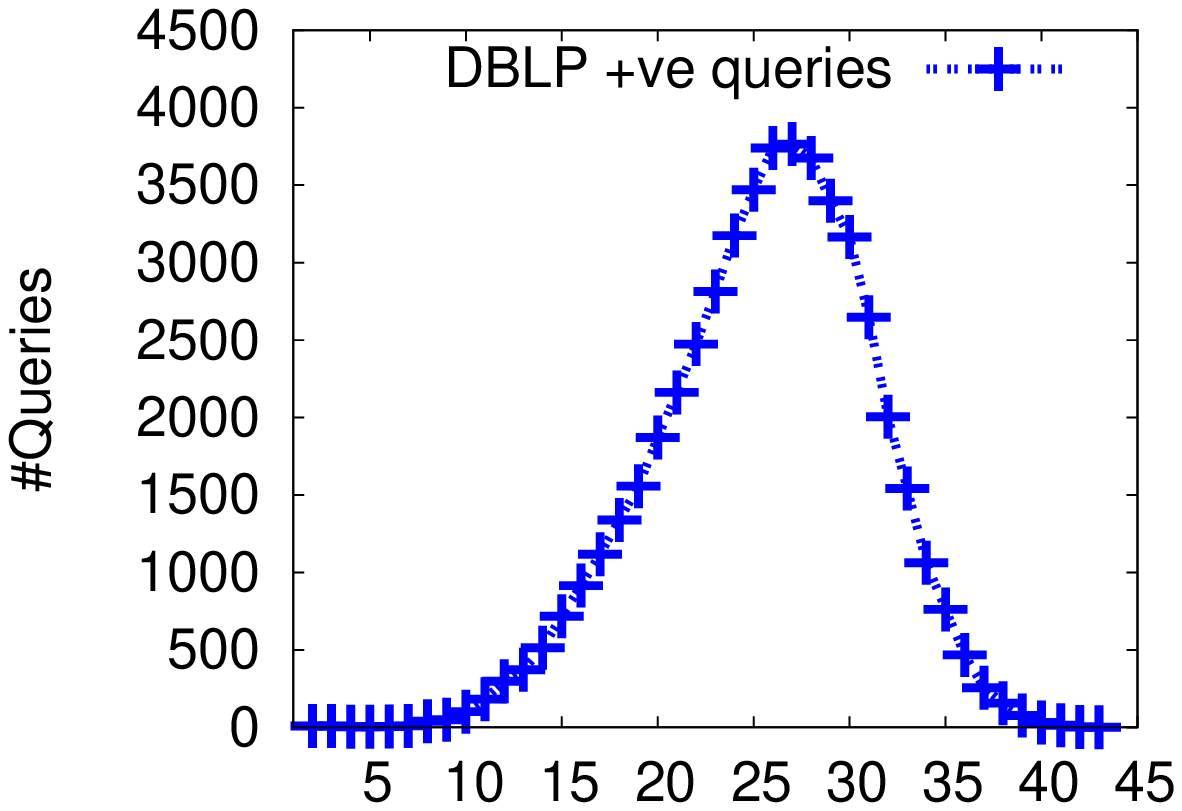}
    }
  }
\vspace{-3mm}
 \centerline{
    \subfigure{
	\includegraphics[scale=0.35]{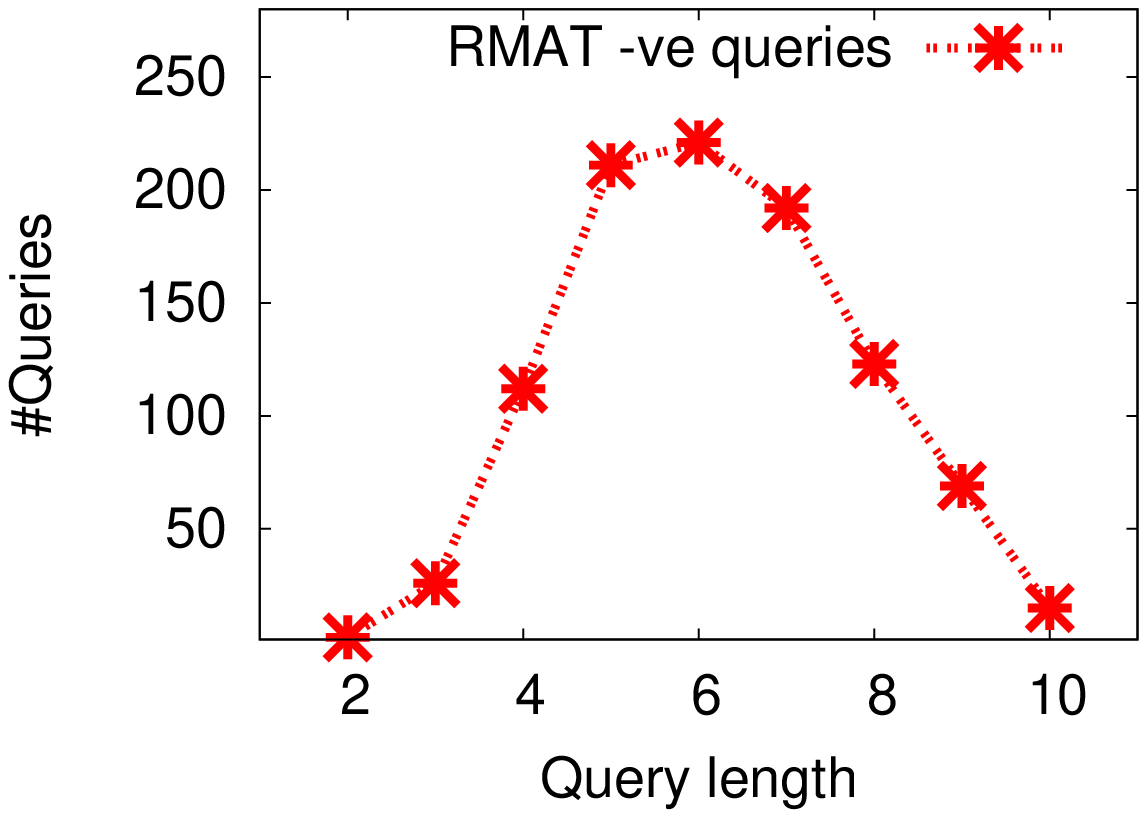}
    }
  \hspace{-0.2in}
    \subfigure{
	\includegraphics[scale=0.35]{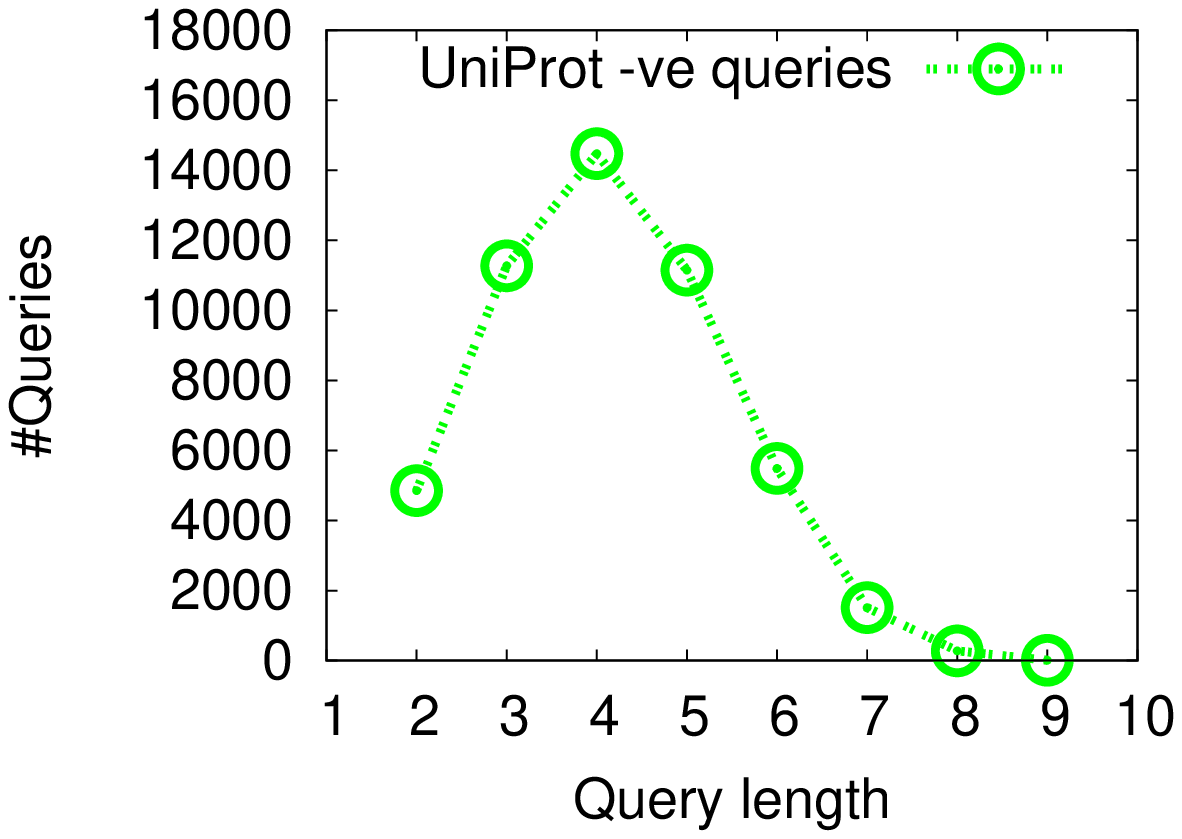}
    }
\hspace{-0.2in}
    \subfigure{
	\includegraphics[scale=0.35]{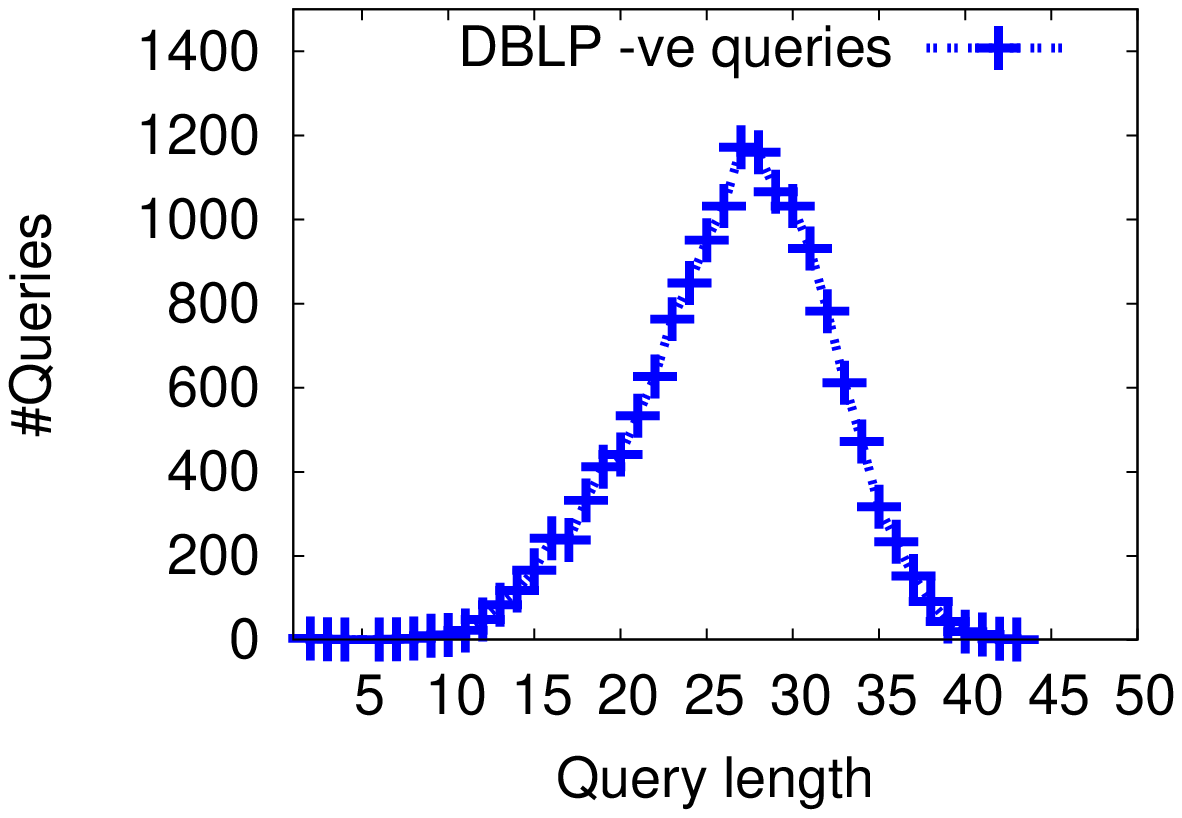}
    }
}
\caption{Distribution of queries as per query-length,
Row 1: Positive queries, Row 2: Negative queries for all datasets.}\label{fig:qcnt_q_size}
\end{figure}

Figure \ref{fig:qcnt_q_size} shows distribution of sampled queries according to their
length. Query length is the number of edge labels appearing in the label order
given in the query.

\end{document}

%% file: intro.tex
\section{Introduction} \label{sec:intro}

With the advent of Semantic Web and RDF, graphs have become richer where
edge labels represent the type of relationship between two nodes which
are connected by that edge.
Exploring paths in the graphs has been a well studied problem. Typically a path
query problem either simply asks for existence of any path between two non-adjacent
nodes in the graph ({{\it reachability} queries), or provides a {\it regular expression}
for the path, and asks for all pairs of nodes that have at least one path between them
which satisfies the given regular expression ({\it regular path queries}).
\textit{Constrained reachability} queries ask  if
the given destination node is reachable from the source node through a path which
satisfies the \textit{constraints} given in the query. The definition of constraints
depends on the targeted problem, e.g., number of hops, type of edge labels, or
a regular expression etc. Constrained reachability queries find their applications
in a variety of graph databases.
Application of label-constrained-reachability (LCR) queries have been outlined by Jin et al.
in \cite{JIN}. In our work, we have focused on a different type of constrained reachability problem
-- \textit{label-order-constrained-reachability} (LOCR). Section \ref{sec:relwork} gives
a broad level overview of the prior work done in these three categories of path queries.

\textbf{Motivating Example:} Consider a graph database of
KEGG\footnote{\scriptsize{Kyoto Encyclopedia of Genes and Genomes http://www.genome.jp/kegg/}} metabolic
pathways.
KEGG dataset has huge directed graphs with nodes as the metabolic compounds, and edge
labels as the enzymes which carry out that metabolic process.
Metabolic pathways, compounds, and enzymes
among different species are not exactly same.
Consider a metabolic path between two compounds (C00267, alpha-D-Glucose and
C00074, Phosphoenolpyruvate) in humans, which is invoked by a series of enzymes (which are
the edge labels on the pathway between the two compounds).
We are interested
in finding -- (1) if an evidence of a pathway exists between those two compounds in some other animal,
say dog -- this is a simple \textit{reachability} query, and (2) if one or more
of the enzymes on the path between C00267 and C00074 in humans also exist on a path in a dog.
Since the pathways may not be
exactly same, we want to know if an enzyme `K12407' appears on the path, followed
by enzyme `K01810' somewhere on the path, followed by enzyme `K00134' etc.
This \textit{constrained reachability} query can be expressed as -- \textit{does there exist
any path from node `C00267' to `C00074', such that the edge labels on the path satisfy a \textit{regular expression}
``*K12407.* K01810.* K00134.*''?}
 Note that expression ``{\it *K12407.* K01810.* K00134.*}'' is written 
according to the Perl regular expression syntax. The `*' before `K12407' allows
zero or more other edge labels (enzymes) to appear on the path before
edge label `K12407'. `K12407 .*' means, at least one edge with label `K12407' should
be present followed by zero or more other edges, before an edge with label
`K01810' appears.
Since we are interested in the \textit{order} in which the edge labels appear, we call these as
\textit{label-order-constrained-reachability } (LOCR) queries.

\begin{figure}
	\centering
		\includegraphics[width=4in,height=1.25in]{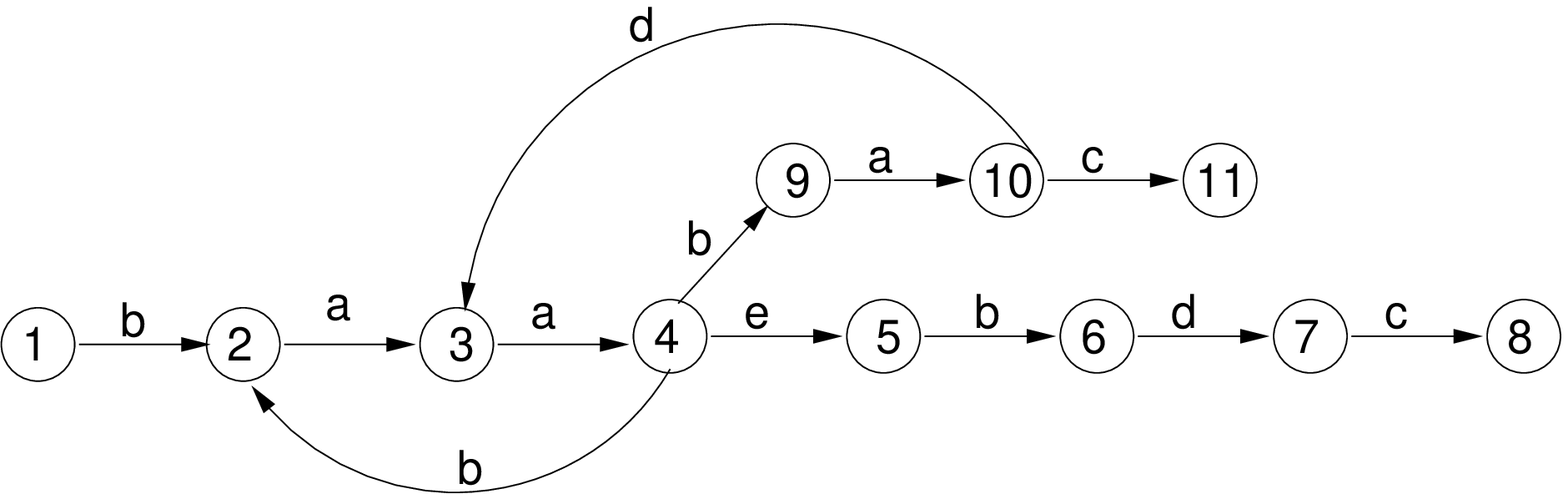}
	\caption{Example Edge-labeled Graph}
	\label{fig:introfig}
\end{figure}

To explain further, consider a sample graph given in Fig. \ref{fig:introfig}.
A regular path
query may ask ``\textit{what all pairs of nodes have at least one path between them which
satisfies edge-label expression ``*a.*b.*c.*''?}''. The answer to this query is
pairs (1, 8), (1, 11), (2, 8), (2, 11), (3, 8), (3, 11), (4, 11), and (4, 8).
Note that (4, 8) pair has a path satisfying ``{\it *a.*b.*c.*}'' due to the cycle among nodes
3, 4, 9, and 10. 
An LOCR query asks -- \textit{does there
exist any path between nodes 3 and 8 such that the edge-labels on the
path satisfy regular expression ``\textit{*a.*b.*c.*}''?} -- answer is Yes.
But answer to another LOCR query --
\textit{does there exist any path between nodes 3 and 8 such that the edge labels
on the path satisfy regular expression ``\textit{*a.*c.*b.*}''?} -- is No.

Path queries and indexing mechanisms have been well studies in the context of
XML data, but each XML document -- represented as an XML graph --
has smaller size
thereby limiting the
number of paths in each graph. For smaller number of nodes and paths, it is
possible to index the paths for path queries.
But for other graphs, specifically like RDF, the number of paths in the graph can be
prohibitively large, which can preclude any path indexing methods. E.g.,
the RDF graph of DBLP data with 13 million edges and 5 million nodes has
more than $10^{25}$ distinct paths.
Unlike previously used path indexing methods,
our BitPath indexing and LOCR query algorithm focuses mainly on building simple but space and time efficient
indexes using \textit{compressed bit-vectors}. Further, we make use of these compact
indexes to \textit{greedily prune} the search space and use a \textit{divide-and-conquer}
algorithm to evaluate the given LOCR query.
Section \ref{eval} has given details about how
compressed bit-vectors can save significant amount of space for indexing large graphs.

Our main contributions in this work are:
\begin{enumerate}
 \item A \textit{light-weight} graph indexing scheme based on compressed
bit-vectors -- \textit{BitPath}.
 \item A \textit{divide-and-conquer} query algorithm using \textit{greedy-pruning} strategy
for  efficient pruning of the search space using the aforementioned indexes.
 \item Experiments on large graphs -- more than 22 million edges and 253
distinct edge labels. To the best of our knowledge these graphs are the largest
amongst the published literature on path queries.
\end{enumerate}

%% file: related.tex
\section{Related Work} \label{sec:relwork}
We have divided the previous literature on path queries into three main categories.

\textbf{Reachability Queries:} Reachability queries just ask if a source node has
\textit{any} path to the destination node.
By definition, the problem of LOCR queries is not same as pure reachability queries and hence
we do not cover the literature of reachability queries here.
For interested reader, some of the recent work by Yildirim et al \cite{HILMI} and Schaik et al \cite{SCHAIK}
on reachability queries give an extensive overview of the previous literature.

\textbf{Regular Path Queries:}
The indexing schemes on graphs for evaluating path queries can be grouped into the following three
categories: \textbf{(1) P-indexes} (path indexes which typically index paths)
-- \textit{Index Fabric} \cite{COOPER}, APEX index \cite{APEX},
building equivalence classes of nodes \cite{MILO,ABITEBOUL2,KAUSHIK,CHEN,MKINDEX} etc.
Path indexing cannot be used for RDF graphs as the number of distinct paths are prohibitively
large.
\textbf{(2) D-indexes} \cite{LI} (node index -- used for determining in constant time
ancestor-descendant relationship). \textbf{(3) T-indexes} (for path-pattern queries such as
\textit{twig} in case of XML) \cite{MPMGJN,StackTree,TwigStack,TwigStackList,Twig2Stack}.
D-indexes and T-indexes can only be
used in the context of XML graphs as it is non-trivial to decide
ancestor-descendant relationship in constant time on a directed graph which does not assume acyclic
structure \cite{HILMI}. 
\textit{Bitmapped Path Index} (BPI) \cite{BPI} and \textit{BitCube} \cite{BitCube} use bit-vector based
indexes. But note that BPI and BitCube use the bitmapped indexes to
index \textbf{paths} in the XML graphs, whereas BitPath uses compressed bit-vectors to only index
the {\em unique edges} in the graph. While a vast number of techniques have been
proposed for path indexing in XML graphs, those approaches cannot be used for
general edge-labeled graphs since XML is widely viewed as tree structured data.

\textbf{Constrained-Reachability Queries:}
Jin et al. \cite{JIN} propose a method for evaluating
label constrained reachability (LCR) queries. \textit{Given a source
and destination node, and a set of edge labels S,
does there exist any path between them, such that each edge label on that path is
in S? No labels other than those in $S$ can appear on that path.}
This problem is different from LOCR queries in the respect that it does not
impose any order among the edge labels, and it does not allow
any edge labels other than the ones in $S$ to appear on the path.
Jin et al. have evaluated their algorithm on synthetic and real graphs
of up to 100,000 nodes and up to 150,000 edges and have compared with DFS
and focused-DFS as the baseline methods.
Fan et al \cite{FAN} have addressed the problem of adding regular expressions and
patterns to the reachability queries. They have evaluated their algorithm on
the graphs of up to 1 million nodes and 4 million edges (synthetic). Gubichev and Neumann
\cite{NEUMANNPATH} have implemented a technique of evaluating path queries over RDF graphs
using purely database style indexing and efficient join processing techniques. Although
Gubichev and Neumann have shown experiments on very large RDF graphs, their queries
use more join-like expressions than rich path-patterns.

%% file: index.tex
\section{BitPath Indexes}  \label{sec:index}
Let $G=(V, E, L, f)$ be a directed edge labeled graph, where $V$ is the set of
nodes, $E$ is the set of edges, $L$ is the set of unique edge labels, and $f$ is
a edge-labeling function $f: E \rightarrow L$, for each edge $(v_i, v_j)$.

Before building the BitPath indexes, we transform the given graph into a
directed acyclic graph by collapsing the strongly connected components (SCCs).
Since the LOCR queries specify constraints on the edge labels on the path, it is
imperative to capture the edges (and their labels) that get collapsed in a SCC.
First, SCCs in a graph are identified using Tarjan's algorithm\footnote{\scriptsize{\url{
http://en.wikipedia.org/wiki/Tarjan's_strongly_connected_components_algorithm}}}.
Let $z$ be a new node to represent the SCC \textit{`C'} after collapsing it. For each edge $e$ in $C$
that gets \textit{removed} as a result of collapsing the SCC, add a
\textit{self-edge} $(z, z)$ with label $f(e)$ in the graph.
The purpose of adding self-edges with same edge-labels is to keep track of the
edge labels that appear in a given SCC. These edges help in determining paths
going through an SCC without having to traverse the entire SCC subgraph at query
time.
\begin{figure*}
	\centering
		\includegraphics[width=5in,height=1.2in]{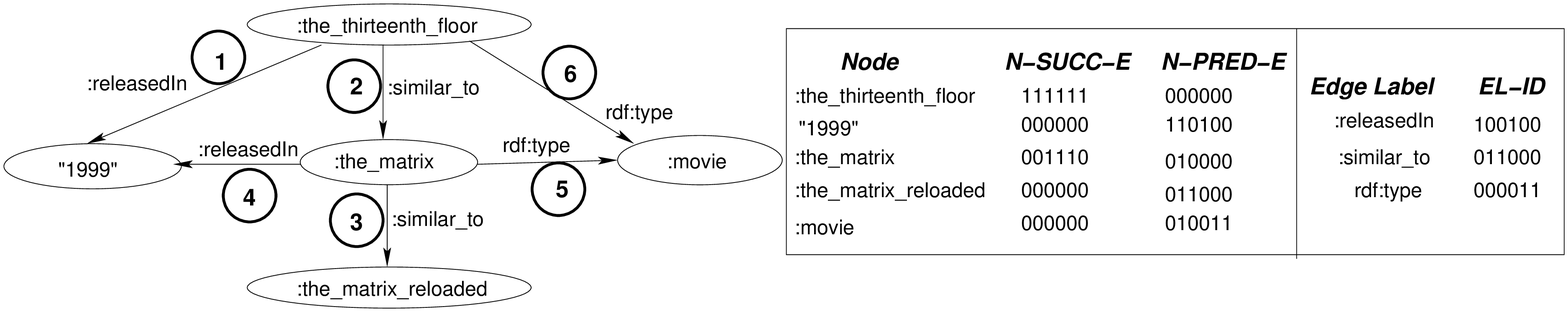}
	\caption{BitPath Indexes using Compressed Bit-vectors}
	\label{fig:example}
\end{figure*}

A label-order-constrained-reachability (LOCR) query requires knowledge of the
\textit{relative order} of edge labels occurring on a given path.
As pointed out in Section \ref{sec:intro} and \ref{sec:relwork} previously, it is
computationally infeasible to index paths of the order of $10^{25}$ or more due to time and space
constraints.
We solve the problem of LOCR queries by creating
four types of indexes on a graph, and using a query answering algorithm,
based on a combination of \textit{greedy-pruning} and
\textit{divide-and-conquer} method.
The four types of indexes are as follows:
\begin{enumerate}
 \item \textit{EID (edge-to-ID):} Each edge $e \in E$ is
mapped to a unique integer ID. For instance, for the graph shown in Fig.
\ref{fig:example}, edge \textit{(:the\_matrix, :movie)} with label
\textit{rdf:type} is mapped to ID 5.
 \item \textit{N-SUCC-E (node's successor edges):} For each node, we index IDs
of all the successor edges, i.e., edges that will get visited if we traverse
the subgraph under the given node. The self edges added to the
graph as a result of collapsing an SCC can be handled trivially by examining the
head and tail of the given edge.
In Fig. \ref{fig:example}, node
\textit{:the\_thirteenth\_floor} will have edge IDs 1, 2, 3, 4, 5, 6 in its
successor list.
\label{indx2}
 \item \textit{N-PRED-E (node's predecessor edges):} Similarly, for each node
we index the predecessor edges, i.e., edges that will get visited if we make
a \textit{backward} traversal on the entire subgraph above the given node.
In Fig. \ref{fig:example}, node \textit{:movie} will have edge IDs 2, 5, 6 in
its predecessor list.
 \item {\it EL-ID (edge label to edge ID):} For each unique edge label $l \in
L$, we index IDs of all the edges in $E$ which have edge label $l$. In Fig.
\ref{fig:example}, edge label \textit{rdf:type} will have IDs 5 and 6 in its
list.
\end{enumerate}

In practice, we use bit-vectors of length $|E|$ (total number of edges in the
graph), for building N-SUCC-E, N-PRED-E, and EL-ID indexes. Each bit position in
the bit-vector corresponds to the unique ID assigned to an edge as per the EID
index.
Fig. \ref{fig:example} shows the indexes for the given example graph.
We apply run-length-encoding on N-SUCC-E and
N-PRED-E indexes of each node depending on the compression ratio.
Typically the unique edge labels in the graph are much fewer compared to the
number of nodes. Hence in an EL-ID index of an edge label, there are many more
interleaved 0s and 1s as compared to an N-SUCC-E or N-PRED-E index, which 
precludes the benefit of run-length-encoding.
Hence we do not apply run-length
encoding on the EL-ID index. Note that at the time of querying we do not
uncompress any compressed indexes. All the algorithms are implemented to perform
bitwise operations on both the gap-compressed indexes as well as non-compressed
indexes.

\subsection{BitPath Index Creation Procedure} 
\label{indexsubsec}

We create EID, N-SUCC-E, and EL-ID indexes in one depth-first-search (DFS) pass over the DAG
generated after collapsing the SCCs as outlined previously.
N-PRED-E index is created by making one backward DFS pass on the graph and using
the EID mapping generated in the first pass.

\textbf{Bit-vector Compression:}
In the first pass, we find all the nodes with 0 in-degree (the root nodes).
For each root node, we make a DFS pass over the entire subgraph
below it.
Starting with ID 1, every new edge encountered in the DFS pass is given a new ID
sequentially. For instance, in Fig. \ref{fig:example}, starting at root node
\textit{:the\_thirteenth\_floor}, we visit edge \textit{(:the\_thirteenth\_floor,
``1999'')} with label \textit{:releasedIn} first and assign ID ``1'' to it. Next when
we visit edge \textit{(:the\_thirteenth\_floor, :the\_matrix)} with label
\textit{:similar\_to} it is given ID ``2''. Since this is a DFS traversal, we
continue traversing the subgraph below node \textit{:the\_matrix}, and
sequentially assign ID ``3'' to next edge visited.
While assigning IDs to the edges, we maintain N-SUCC-E list for each
node encountered on the DFS path as well as a \textit{stack} of DFS nodes
visited on the given path.
When an ``unvisited'' node is encountered, we add all the outgoing edges of that node in
the N-SUCC-E list of each node in the DFS stack.
Once a node is marked ``visited'', we build a bit-vector of  N-SUCC-E list.
Note that we have already collapsed any strongly connected components in
the graph, hence the current graph is acyclic.
If a ``visited'' node is encountered again in the DFS walk, we simply
add all the edge IDs corresponding to the 1-bit positions in
its N-SUCC-E bit-vector to the N-SUCC-E list of all the nodes
in the DFS stack.

Since IDs to the edges are assigned  as they are visited,
this scheme generates N-SUCC-E bit-vectors with \textit{large
gaps} for most nodes.
A bit-vector with large-gaps is the one where a lot of 0s or 1s appear together, e.g.,
a bit-vector ``111111000001111111'' has large-gaps as opposed ``11001010101010'' which
has small gaps.
We make use of this fact to apply \textit{run-length encoding} on N-SUCC-E bit-vectors.
Note that this was a heuristic observed for many real-life graphs, and it is
possible to generate a pathological graph where run-length encoding does not
fetch the desired benefit.
Since for a typical RDF or any edge-labeled graph, $|L|$ is much smaller than
$|V|$, EL-ID index is typically smaller than EID and N-SUCC-E indexes (and
even N-PRED-E discussed below).
At the end of the first DFS pass, we get EID, N-SUCC-E, and EL-ID indexes.

In the second pass, we start from the leaf nodes -- nodes with 0 out-degree --
and make a \textit{backward DFS} traversal on the graph. N-PRED-E index for each node is
built in the same way as the N-SUCC-E index. The only difference is that in the
second pass, we utilize the EID index that was populated in the first pass.
Hence every time we encounter an
edge, we simply look up its ID in the EID index and use it to construct the
N-PRED-E index. When a node is marked ``backward-visited'' in this pass, we
generate its bit-vector N-PRED-E index in the same manner as N-SUCC-E index.
At the end of the second pass, all the indexes are written to the disk.

%% file: query_algo.tex
\section{BitPath Query Algorithm} \label{algo}
In an LOCR query \textit{(x, y, (a.*b.*c.*....l.*))} -- we want to find if there exists
any path between source $x$ and destination $y$, such that labels
\textit{(a, b, c,...l)} appear on that path in the given order (other edge
labels can appear on this path as well, ref. Section \ref{sec:intro}).
The ``...'' denotes that
there can be any number of labels specified in the order-constraint.
\begin{figure}
	\centering
		\includegraphics[width=4in,height=1.2in]{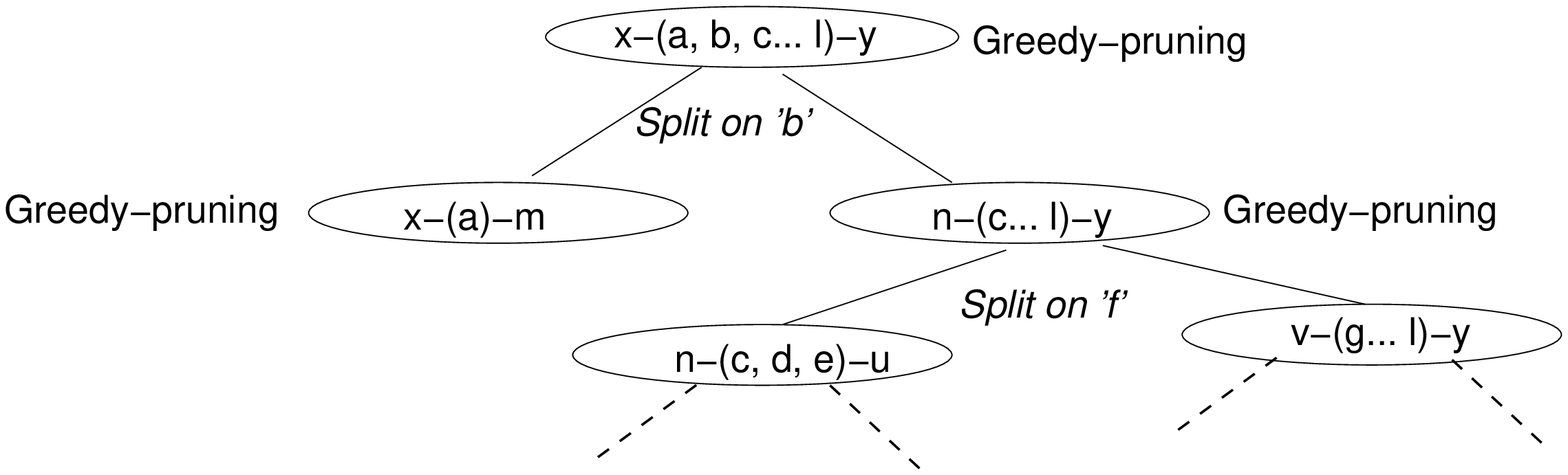}
	\caption{Divide-and-conquer with Greedy pruning}
	\label{fig:divideconq}
\end{figure}

The evaluation of an LOCR query can be broken down into following steps:
\begin{enumerate}
 \item Is $y$ is reachable from $x$?
 \item If yes, we want to find if there exists
any path between $x$ and $y$, such that all of the labels $a$, $b$, $c$ ... $l$
appear somewhere on that path in that order.
 \item Suppose there exist more than one paths where label $b$ appears
somewhere on the path. We find an edge $(m, n)$ with label $b$ on one of those
paths. \label{step1}
 \item Next, we find if there exists any path between $x$ and $m$, such
that label $a$ appears somewhere on it. If not, go back to Step \ref{step1} and choose
another edge $(s, t)$ with label $b$ on another path between $x$ and $y$. \label{step2}
 \item Next we find if there exists a path between $n$ and $y$, such that labels $c$ ... $l$ appear
somewhere on it in the given order. If not, go back to Step \ref{step1}, and choose
another edge $(p, q)$ with label $b$ on a path between $x$ and $y$ and so on.\label{step3}
 \item We recursively go through above steps by dividing the query into smaller sub-queries, until 
the available search space is exhausted or a satisfying path is found.
\end{enumerate}

Now let us see how we can use the BitPath indexes to carry out above steps
in an efficient manner.
N-SUCC-E index of a node gives us all the edges that can be reached from
that node and N-PRED-E index gives the edges that can eventually
lead to a node. If the intersection of N-SUCC-E index of node $x$ and
N-PRED-E index of node $y$ is non-empty, node $y$ is reachable from $x$. This answers the first
step above.
If $\{N\mbox{-}SUCC\mbox{-}E(x) \cap N\mbox{-}PRED\mbox{-}E(y) \cap
EL\mbox{-}ID(b)\} \neq \phi$, i.e., the intersection of successor index of $x$,
predecessor index of $y$ and EL-ID index of label $b$ is non-empty, there is at
least one path from $x$ to $y$, where edge label $b$ appears somewhere on the
path. This solves our second step. N-SUCC-E($x$), N-PRED-E($y$), and EL-ID($b$)
are bit-vectors and their intersection requires two bitwise AND operations.
Let the result of this bitwise AND operation be \textit{INTSECT}.

Position of a 1 bit in INTSECT bit-vector gives EID of an edge with label $b$.
A reverse look up in the EID index gives us an edge, say $(m, n)$, with label
$b$. This satisfies the third step above. If $\{N\mbox{-}SUCC\mbox{-}E(x) \cap
N\mbox{-}PRED\mbox{-}E(m) \cap EL\mbox{-}ID(a)\} \neq \phi$, it means that there
exists at least one path between $x$ and $m$ where edge label $a$ appears
somewhere on the path. This addresses the forth step. If we put the earlier
result and this result together, it tells us that there exists at least one path
between $x$ and $y$, such that edge label $a$ appears somewhere before label
$b$. Recursively, we solve our fifth step to find if there exists any path between
$n$ and $y$ such that edge labels $c...l$ appear somewhere on it. This procedure is
depicted in Fig. \ref{fig:divideconq}.
This is a divide-and-conquer binary tree.
Note that any subquery (node) in this tree,
which has only one edge label in its sequence is a \textit{leaf-node}.
The total number of nodes in this tree are $2*|\{a,b,c,...l\}| - 1$, where $|\{a,b,c,...l\}|$ denotes
the total number of edge labels in the LOCR query.
The query processing stops when either all
the edges at every subquery in the tree are exhausted, or if a ``Yes'' answer is found at all the
leaf nodes -- which is propagated back to their respective parents recursively.

In the example above, we chose to split the edge label order ``\textit{a, b, c,...l}''
on label $b$ first for the ease of understanding. But for further optimization, we can choose the
\textit{split-point} depending on the \textit{selectivity} of the intersection
of N-SUCC-E($x$), N-PRED-E($y$), and EL-ID for each label in the query\footnote{\scriptsize{Selectivity is inversely proportional
to the number of edges in the intersection -- high selectivity means fewer edges and low selectivity
means more edges.}}. We will always choose to split on the edge label with high selectivity.
This is the \textit{greedy-pruning} step, which is outlined in Algorithm \ref{alg:greedy}.

\begin{algorithm}
\begin{algorithmic}[1]
\scriptsize{
\caption{greedy\_pruning(\textit{x, y, label\_seq})} \label{alg:greedy}
\STATE min\_label = 0
\STATE $min\_edges = \infty$
\FOREACH{$l$ in \textit{label\_seq}}  \label{ln:minint}
\STATE $intsect = N\mbox{-}SUCC\mbox{-}E(x) \cap N\mbox{-}PRED\mbox{-}E(y) \cap EL\mbox{-}ID(l)$
\IF{$|intsect| < |min\_edges|$}
\STATE $min\_edges = intsect$
\STATE min\_label = $l$
\ENDIF
\ENDFOR \label{ln:minint-end}
\RETURN pair($min\_edges$, min\_label) \label{ln:pairret}
}
\end{algorithmic}
\end{algorithm}

In Algorithm \ref{alg:greedy}, \textit{label\_seq} is the order of edge labels in the query.
If at any given sub-query node in the divide-and-conquer tree (see Fig.
\ref{fig:divideconq}) there are more than one edge labels in the given
\textit{label\_seq}, the \textit{greedy-pruning} strategy takes intersection of
$(N\mbox{-}SUCC\mbox{-}E(x)\ \cap\ N\mbox{-}PRED\mbox{-}E(y)\ \cap
EL\mbox{-}ID(l))$ for each label $l$ (lines \ref{ln:minint}--\ref{ln:minint-end}).
The edge label  which generates the smallest edge intersection
set is denoted as \textit{min\_label}, and the set of edges as \textit{min\_edges}.
The \textit{greedy-pruning} method returns \textit{min\_label} and \textit{min\_edges} (Line \ref{ln:pairret})
which are subsequently used to partition the initial query into two sub-queries.

For typical real life graphs like RDF, there are few distinct edge labels ($L$
is very small) as compared to the total number of edges $E$. For instance, the UniProt RDF
graph of 22 million edges has only 91 distinct edge labels. Moreover, the distribution of
these edge labels is not uniform, a large number of edges have few distinct edge
labels. In the UniProt dataset, the edge label \textit{``rdf:type''}
appears on $\sim$5 million edges, about 10 edge labels appear on 1 million edges
each and about 20 edge labels occupy $\sim$100,000-200,000 edges each. We
exploit this skewed label distribution to effectively prune the potentially
large search space of edges.
The skewed edge distribution holds true for most real life graphs, but it
is possible to synthetically build graphs where there is one root node, one sink
node, and a set of edge labels that follow uniform distribution.
For such a graph, given the root and sink nodes and a list of edge labels in an LOCR query, the
first step of \textit{greedy\_selection} will not be able to achieve any pruning because every edge
will be in the N-SUCC-E index of source node and N-PRED-E index of the sink
node. However, as the divide-and-conquer tree grows deeper, further sub-queries can 
exploit the BitPath indexes to prune the search space, which will preclude exploring
all the paths.
Algorithm \ref{alg:divconq} describes the \textit{divide-and-conquer} strategy.

\begin{algorithm}
\begin{algorithmic}[1]
\scriptsize{
\caption{divide-and-conquer(\textit{x, y, label\_seq})} \label{alg:divconq}
\STATE res = FAIL
\IF{topo\_order[$y$] - topo\_order[$x$] $<$ label\_seq.size()} \label{ln:topoorder}
\STATE return FAIL
\ENDIF
\STATE pair(min\_edges, min\_label) = greedy\_pruning($x$, $y$, label\_seq)\label{ln:greedy}
\STATE
\IF{min\_edges == $\phi$}
\RETURN FAIL \label{ln:empty}
\ENDIF
\IF{label\_seq.size() $<=$ 1}
\RETURN SUCCESS
\ENDIF
\STATE
\STATE lseq1 = get\_seq(label\_seq.begin(), min\_label-1); \label{ln:splt1}
\STATE lseq2 = get\_seq(min\_label+1, label\_seq.end()); \label{ln:splt2}
\STATE
\FOREACH{$eid$ in min\_edges} \label{ln:foreachedge}
\STATE edge $e$ = eid\_to\_edge($eid$)\label{ln:eidlook}
\COMMENT{for edge \textit{(k,l)}, $k$ is the tail, and $l$ is the head of the edge}
\STATE res = divide-and-conquer($x$, $e.tail$, lseq1)  \label{ln:recurs3}
\IF{res == SUCCESS}
\STATE res = divide-and-conquer($e.head$, $y$, lseq2)
\IF{res == SUCCESS}
\STATE break
\ENDIF
\ENDIF\label{ln:recurs3-end}

\ENDFOR \label{ln:endforeachedge}
\RETURN res
}
\end{algorithmic}
\end{algorithm}

For the sake of illustration, let us assume an LOCR query over nodes $x$ and $y$ and
a label order \textit{(s, t, u, v, w)}. To evaluate this query, first Algorithm \ref{alg:divconq} is
called on $x$ and $y$ nodes with \textit{label\_seq} containing all the edge
labels\textit{(s, t, u, v, w)}. Although this example considers a sequence for
length five, the algorithm is invariant to the length of the query.
Also, an edge label can be repeated any number of times in the sequence,
e.g., \textit{(a, b, a, c, a,)} or \textit{(a, a, a, a, b, b)}.
The \textit{label\_seq} can as well be \textit{empty}, in which case it simply
boils down to a reachability query.

The topological order of nodes gives a hint about the longest incoming
path to any node.
If the difference between the topological order of $x$ and $y$ is lesser than
the length of \textit{label\_seq}, it means that there is no
path from $x$ to $y$ of length $|label\_seq|$ or more.
\textit{Divide-and-conquer} uses this property to preclude 
exploring paths shorter than $|label\_seq|$ (Line \ref{ln:topoorder}).
Using \textit{greedy\_pruning} (line \ref{ln:greedy}) we first get the minimal set of EIDs
(\textit{min\_edges}) such that they are common between the successor edges of
$x$, predecessor edges of $y$, and also contain a label \textit{min\_label}
$\in$ \textit{label\_seq}.

Let \textit{min\_label} be $u$.
Each edge $e$ in \textit{min\_edges}, has same edge label $u$ on it.
Let $e$ be an edge over nodes $(k, l)$ such that
$e.tail = k$, $e.head = l$ (line \ref{ln:eidlook}).
Now the original query is divided into 2 parts -- (1) \textit{is there any path
from node $x$ to $e.tail$ such that it satisfies a label order \textit{(s,
t)}?}, (2) \textit{is there any path from node $e.head$ to $y$ such that it
satisfies a label order \textit{(v, w)}}? If the \textit{label\_seq} is split over
$s$ or $w$, one of the sub-queries will have an empty \textit{label\_seq}.
A sub-query with empty \textit{label\_seq} is just a reachability query.
Such sub-query is skipped as the reachability of the nodes is previously
evaluated in the \textit{greedy-pruning} step.
If the minimum set of edges for any label in \textit{label\_seq} is empty, it
clearly implies that nodes $x$ and $y$ do not have any path with label $l$
between them.
In this case, we stop exploring the path further and return (line
\ref{ln:empty}).

\textbf{Asymptotic Analysis:}
With respect to the runtime complexity, the in-memory program stack of the \textit{divide-and-conquer}
algorithm is at most the size of original label
sequence of the query. That is because this is the size of the divide-and-conquer query tree
as shown in Fig. \ref{fig:divideconq}.
\begin{figure}
	\centering
		\includegraphics[width=4.8in,height=2in]{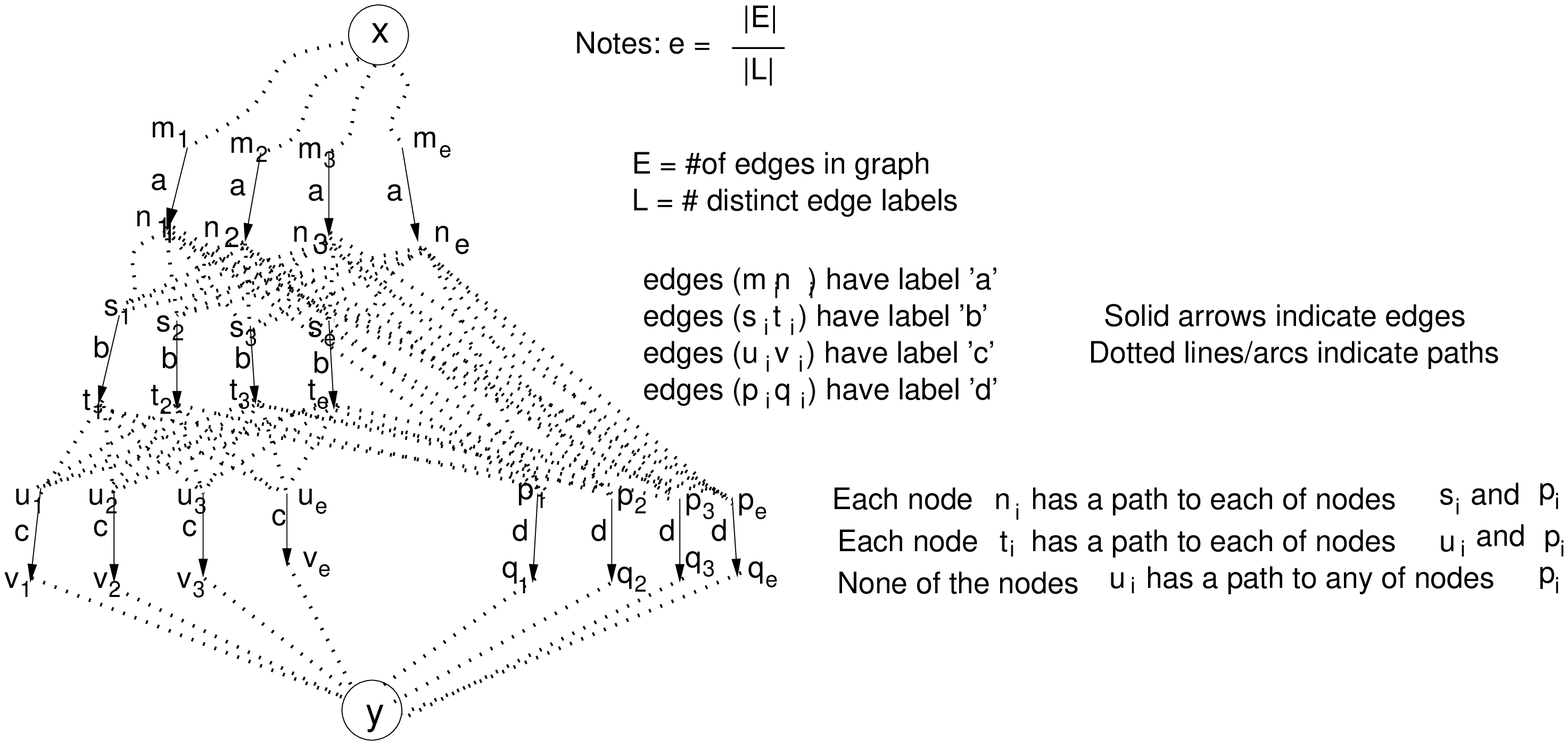}
	\caption{Pathological Example for Exponential Behavior}
	\label{fig:worstcomplex}
\end{figure}

If we assume a uniform distribution of edge labels in the graph,
in the worst case the \textit{divide-and-conquer} algorithm can get called as many as
$(2*|E|/|L|)$ times (Lines \ref{ln:foreachedge}--\ref{ln:endforeachedge} in Algorithm(\ref{alg:divconq}))
for a given call to the \textit{divide-and-conquer} method.
The worst case complexity of the entire algorithm is $O((|E|/|L|)^{|label\_seq|})$
considering the recursive calls to the \textit{for-loop} (Line \ref{ln:foreachedge}).
But this worst case complexity will arise only when the graph and the paths have
a structure as shown Fig. \ref{fig:worstcomplex}.
Fig. \ref{fig:worstcomplex} exemplifies an LOCR query over nodes $x$ and $y$ and \textit{label\_seq} is $(a,b,c,d)$.
The dotted lines denote a \textit{path} and solid lines denote an edge.
In this graph, all nodes of type $n_i$ ($1 \leq i \leq e$, $e=|E|/|L|$)
have a path to all nodes $s_i$ and $p_i$. All $t_i$ nodes have a path to nodes $u_i$ as well as to
$p_i$. Since we do not index paths in the graph, the algorithm cannot know beforehand
if edges with labels $c$ and $d$ in the successor list of nodes $n_i$ or $t_i$ are on the same path.
As shown in the figure, it is possible to generate such a graph synthetically where
the worst case exponential behavior can be observed.
But for all practical purposes, the runtime complexity of the algorithm
is not $O((|E|/|L|)^{|label\_seq|})$ due to the skewed distribution of edge labels
in the real world graphs.

The complexity of \textit{greedy-pruning} is $O(|label\_seq|*|E|)$. But since we use bit-vectors
for storing N-SUCC-E, N-PRED-E, and EL-ID indexes, for all practical purposes
\textit{greedy-pruning} does not imply traversing the entire graph $|label\_seq|$
times.

\textbf{Handling Nodes and Paths in Strongly Connected Components:}
Since every node in an SCC can reach every other node in the same SCC,
it is difficult to decide the start and end of a path going through SCC and 
the order of the edge-labels on that path -- which is required in LOCR query processing.
Hence presently we process the paths going through SCCs by
simply checking the self-edges introduced while merging the SCCs (ref. Section
\ref{sec:index}).

Let an LOCR query be \textit{(x, y, (*a.*b.*c.*d.*e.*))}, such that $x$ and $y$
are part of the same SCC in the original graph. They are represented by node $z$
in the graph obtained by merging SCCs. For such a query, we simply
check if there exist 5 self-edges $(z, z)$ with labels $a$, $b$, $c$, $d$, and $e$.

We use this same technique for BitPath as well as the baseline methods used for
performance evaluation, hence for all practical purposes we have
sampled queries from the directed acyclic graph (with self-edges)
obtained after merging the SCCs for the experimental evaluation.

%% file: eval.tex
\section{Evaluation} \label{eval}

BitPath indexing and query processing algorithm is developed in C and compiled
using g++ (v4.4) with -O3 optimization flag. We used an OpenSUSE 11.2 machine
with Intel Xenon X5650 2.67GHz processor, 48 GB RAM with 64 bit
Linux kernel 2.6.31.5 for our experiments. Although we used a
desktop with 48 GB memory, the BitPath index size for the datasets is much
smaller than that (refer Section \ref{sec:indxsize}).

\subsection{Competitive Methods} \label{sec:compmethods}
Since every RDF graph can be expressed in RDF/XML format, we explored the options of evaluating
LOCR queries on XML representation of an RDF graph, but concluded that an LOCR query cannot
be faithfully translated into an XPath query.
As outlined in Section \ref{sec:relwork}, path indexing approaches suggested in the context
of XML/XPath query processing cannot be used to evaluate LOCR queries on general graphs.
RDF graphs can be represented in XML format\footnote{\scriptsize{\url{http://www.w3.org/TR/rdf-syntax-grammar/}}},
hence we explored the options of evaluating LOCR
This requires translating the given LOCR query into equivalent XML path query.
A faithful translation of an LOCR query into equivalent query on the XML
graph of RDF does not represent a path query. It often has to be processed using
iterative join of two or more tree patterns (twig) queries. Native XQuery
specifications do not support recursive joins of tree pattern queries, where the number
of recursions are not known at the query time. An example of this is given in Appendix-\ref{apdx:rdfxml}.
Also the BitPath method of indexing and processing LOCR queries can be applied to any
other edge-labeled directed graph. But any edge-labeled directed graph -- which does not satisfy RDF
constraints -- cannot be represented as an XML graph.
Hence for competitive evaluation, we used optimized versions of DFS and BFS as the baseline methods.

\textbf{1. Optimized DFS (DFS):} Given an LOCR query between nodes $x$
and $y$, if the out-degree of $x$ is lesser than in-degree of $y$, we start the DFS walk
from $x$ and continue until $y$ is reached and the given path satisfies LOCR \textit{label\_seq}.
If $y$'s in-degree is lesser, we start a \textit{reverse}-DFS walk from $y$ with
\textit{reversed} order of labels and continue up to $x$.
This method is referred to as ``DFS'' in the rest of the text.

\textbf{2. Optimized-Focused DFS (F-DFS):} This method is same as optimized DFS,
but additionally at every node we check the reachability of the destination node
$y$, (or reachability from node $x$ if we perform reverse DFS),
using the intersection of N-SUCC-E and N-PRED-E
BitPath indexes. If $y$ is not reachable from the given node $n$,
$N\mbox{-}SUCC\mbox{-}E(n)\ \cap\ N\mbox{-}PRED\mbox{-}E(y) = \phi$.
This method is further enhanced as follows: Maintain a
\texttt{reachability} array. 
The very first time node $n$ is explored,
update \texttt{reachability[n]} to note if $y$ is reachable from $n$.
If node $n$ is visited again through a different path, next
time just look up \texttt{reachability[n]} to decide if the paths below $n$
should be explored or discontinued.

\textbf{3. Optimized-Bidirectional-BFS (B-BFS):} In bidirectional BFS, we traverse
down the subgraph below $x$ and \textit{above} $y$ one step at a time,
maintaining \textit{forward} and \textit{reverse} BFS queues. Each node enqueued
in the BFS queue has its own $label\_seq_n$ associated with it, which tells
which labels in the original sequence have been seen on a path to node $n$.

We further \textit{optimize} bidirectional-BFS as follows: if a node in the
BFS queue is reached through another ``better'' path (a path
is ``better'' if it has seen more labels in the LOCR {\it label\_seq}, before being
taken out of the BFS queue, that node's $label\_seq_n$ is updated with the
{\em label\_seq} seen on the ``better'' path. In our experience, the optimized
bidirectional BFS performed better than the naive bidirectional BFS by an order
of magnitude.  For simplicity the optimized-bidirectional-BFS method is referred
to as ``B-BFS'' in the rest of the text.

\subsection{Datasets and Queries}

We used 2 real RDF datasets: RDFized DBLP dataset by
LSDIS lab -- SwetoDBLP \cite{SWETODBLP} and a smaller
subset of UniProt protein dataset \cite{UNI}.
We also used a synthetically generated dataset using the R-Mat algorithm \cite{RMAT,GTGRAPH}.
R-Mat graph is assigned edge labels using Zipf \cite{ZIPF}
distribution with distribution parameter ($s$) as 2.95. The statistical
characteristics of these datasets are given in Table \ref{tbl:datasets}.
\textit{``Largest depth''} is the largest depth of a node in the topological
sort. Note that edge and node counts given here are after
collapsing the strongly connected components. The original edge and node counts
are only slightly higher. These graphs are much larger as compared to the previous
published results related to path query processing.

\begin{table*}[!ht]
\centering
\scriptsize{
\caption{Datasets Characteristics}
\label{tbl:datasets}
\begin{tabular}{|c|c|c|c|c|c|c|p{1cm}|c|}
\hline
 & \#Edges & \#Nodes & \#Edge labels & Max-indeg & Max-outdeg & Avg-in/outdeg & Largest depth & SCCs\\
\hline
\hline
R-Mat & 14,951,226 & 4,085,180 & 253 & 19 & 13,053 & 3.65 & 12 & 0\\
\hline
UniProt & 22,589,927 & 6,634,185 & 91 & 800,127 & 1046 & 3.35 & 10 & 423\\
\hline
SwetoDBLP & 13,378,152 & 5,458,220 & 145 & 907,731 & 9245 & 2.44 & 66 & 146\\
\hline
\end{tabular}
}
\end{table*}

We generated two types of queries -- (1) those which have a path satisfying the
given LOCR query -- \textit{positive} queries, (2) those which
do not have any path satisfying the LOCR query but the destination node is
reachable from the source node -- \textit{negative} queries.

In all we generated 50K positive and 50K negative queries on the three datasets using the following procedure:
First, 100K paths are generated through a ``backward'' traversal over the directed
acyclic graph, starting from the leaf nodes. At each node during the backward traversal,
a parent node is selected with either 1) uniform probability, or 2) probability
proportional to the topological order of the node. The second
strategy is incorporated specifically to discover longer paths in the graph.
Subsequently, these paths are used to generate both positive and negative queries.

\begin{figure}[h!]
	\centering
		\includegraphics[scale=0.75]{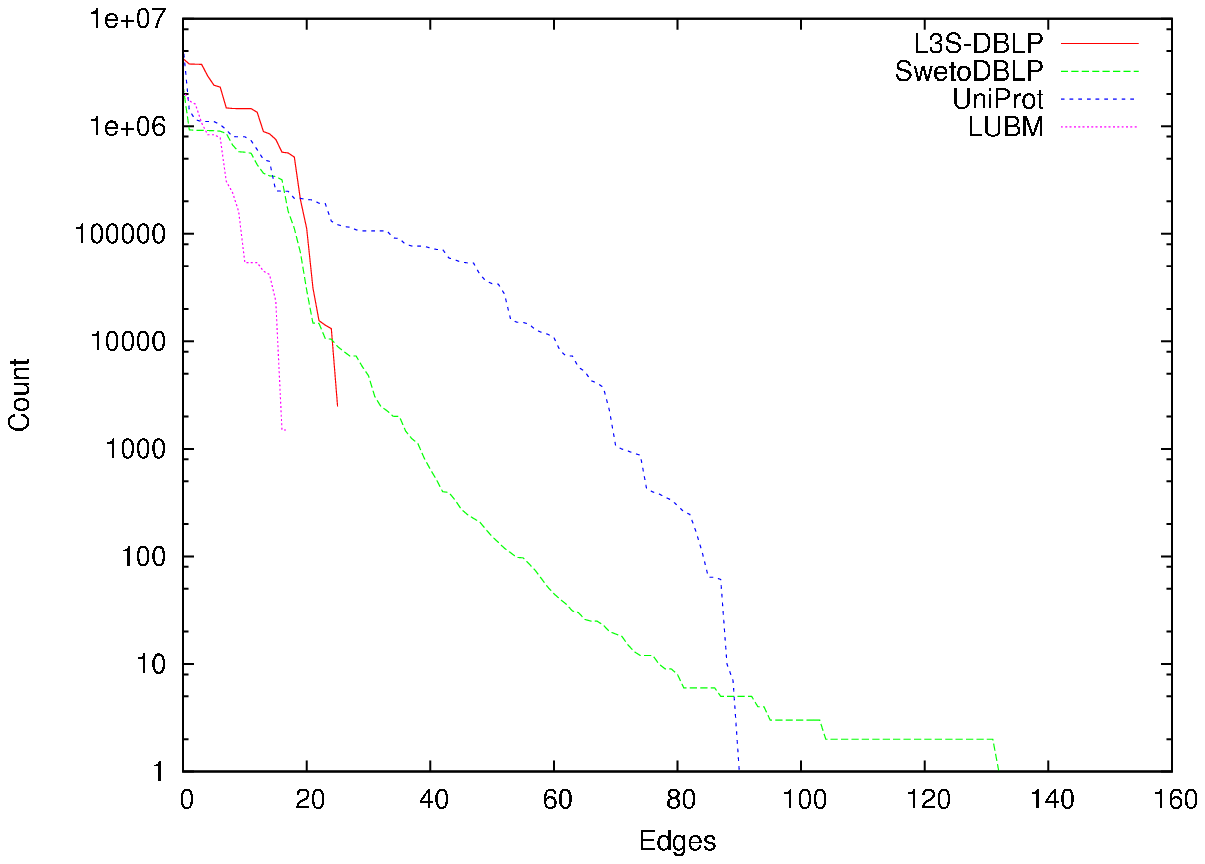}
	\caption{Edge Label Frequencies}
	\label{fig:edgedistr}
\end{figure}

For positive queries, each edge label appearing on a path is removed with
uniform probability. To generate negative queries, we introduce
an extra edge label which does not appear on the given path and
\textit{shuffle} the edge labels to generate a random order. Note that this
method can still result in incorrect negative queries. Without the knowledge of
all paths between a pair of nodes, it is possible that a randomly generated
negative LOCR query might be satisfied by some other path between the same pair
of nodes. In such cases, we simply discard the query.
As outlined previously in Section \ref{algo}, most real-life graphs follow a non-uniform
the edge-label distribution, which benefits in the evaluation of LOCR queries using the greedy-pruning
strategy. The plots of edge-label frequencies for each dataset are given in Fig. \ref{fig:edgedistr}.

\textbf{Evaluation Metrics:}
\textbf{(1)} Average time to run queries of the same query-length, i.e., we grouped
queries of same query-length, and computed average query processing time and
standard deviation for those queries. Length of the query is $|label\_seq|$, e.g., an LOCR query
\textit{(x, y, (*a.*a.*b.*b.*c.*d.*))} has query-length 6.
Since we chose each label on randomly generated paths with uniform probability,
typically an LOCR query of length 6 was generated from a randomly generated path
of length 12. Observe that the query length does not indicate the length of the
\textit{actual path} in the graph which satisfies that query. 
\textbf{(2)} BitPath index construction time for each dataset.
\textbf{(3)} Cumulative size of BitPath indexes for each dataset.

Since we generated the queries with random walks, for the interested reader
we have given the distribution of queries by query length in Appendix-\ref{apdx:qcnt_q_size}.

\subsection{Query Performance}

\begin{figure}[!ht]
  \centerline{
    \subfigure[]{
      \label{fig:rmat_bitpath}
		  \includegraphics[scale=0.43]{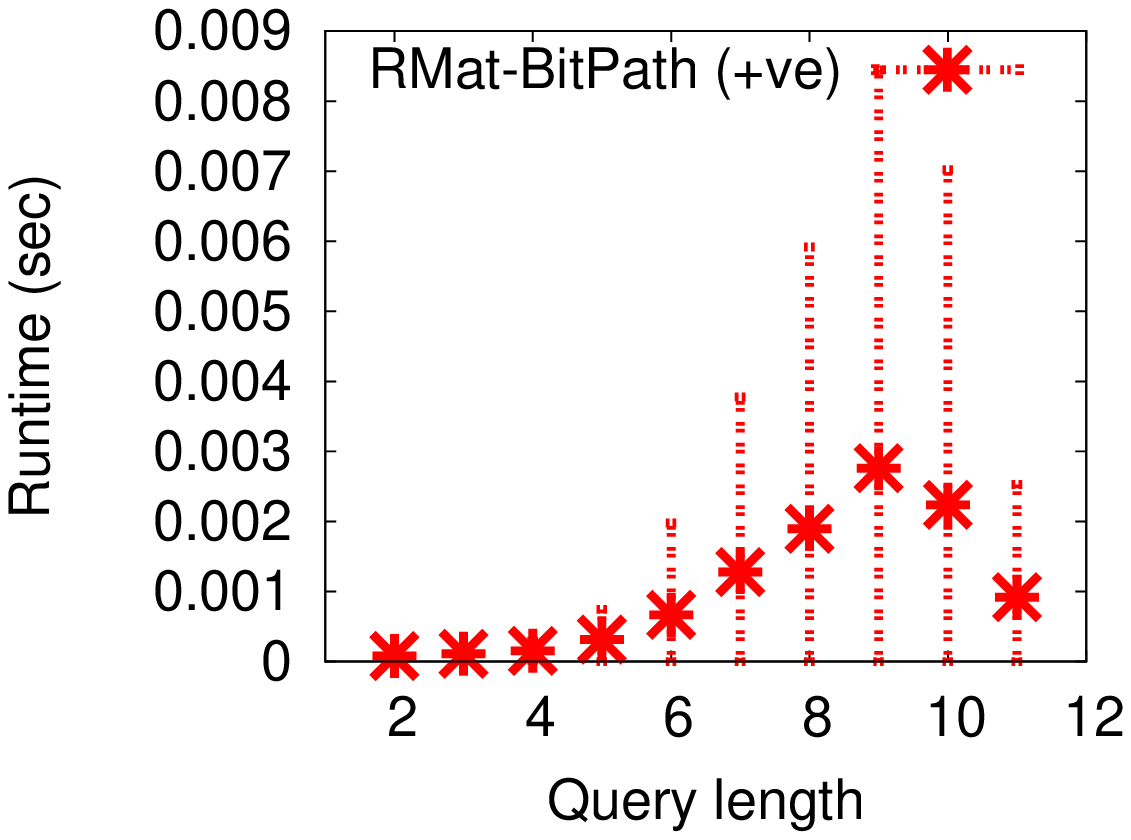}
    }
    \subfigure[]{
      \label{fig:rmat_neg_bitpath}
		  \includegraphics[scale=0.43]{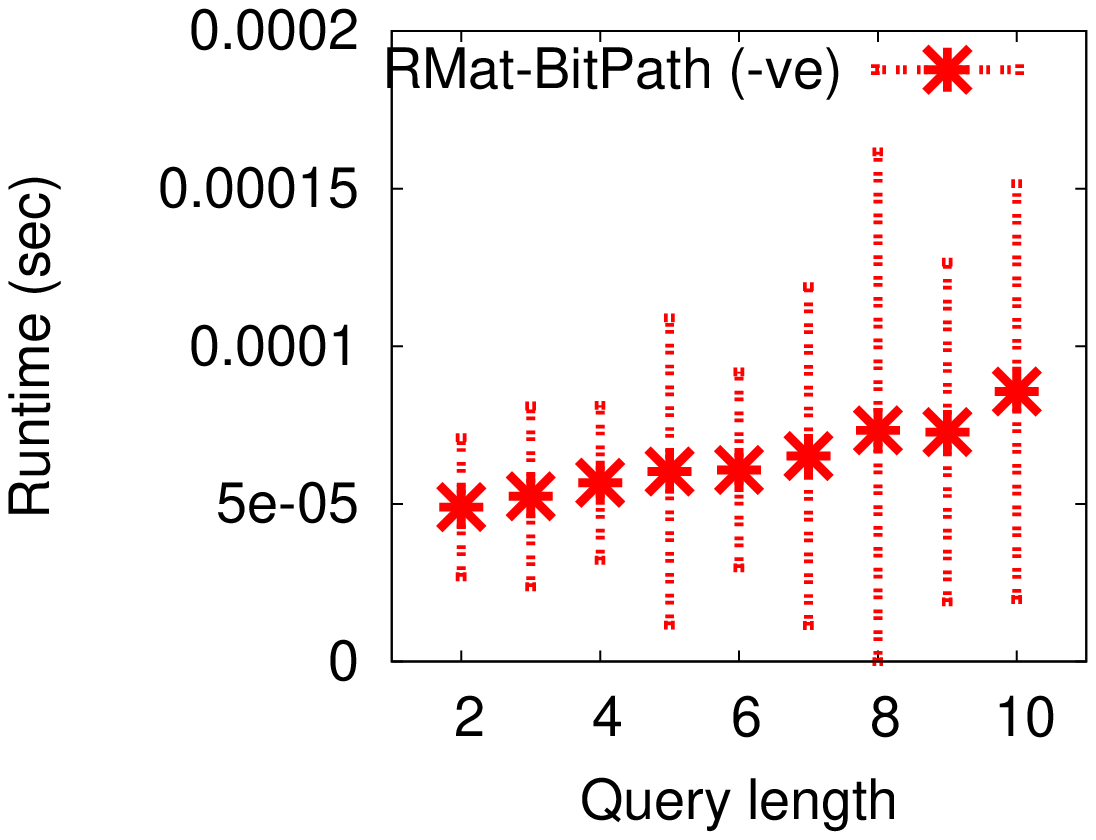}
    }
  }
  \centerline{
    \subfigure[]{
      \label{fig:rmat_bfs}
		  \includegraphics[scale=0.43]{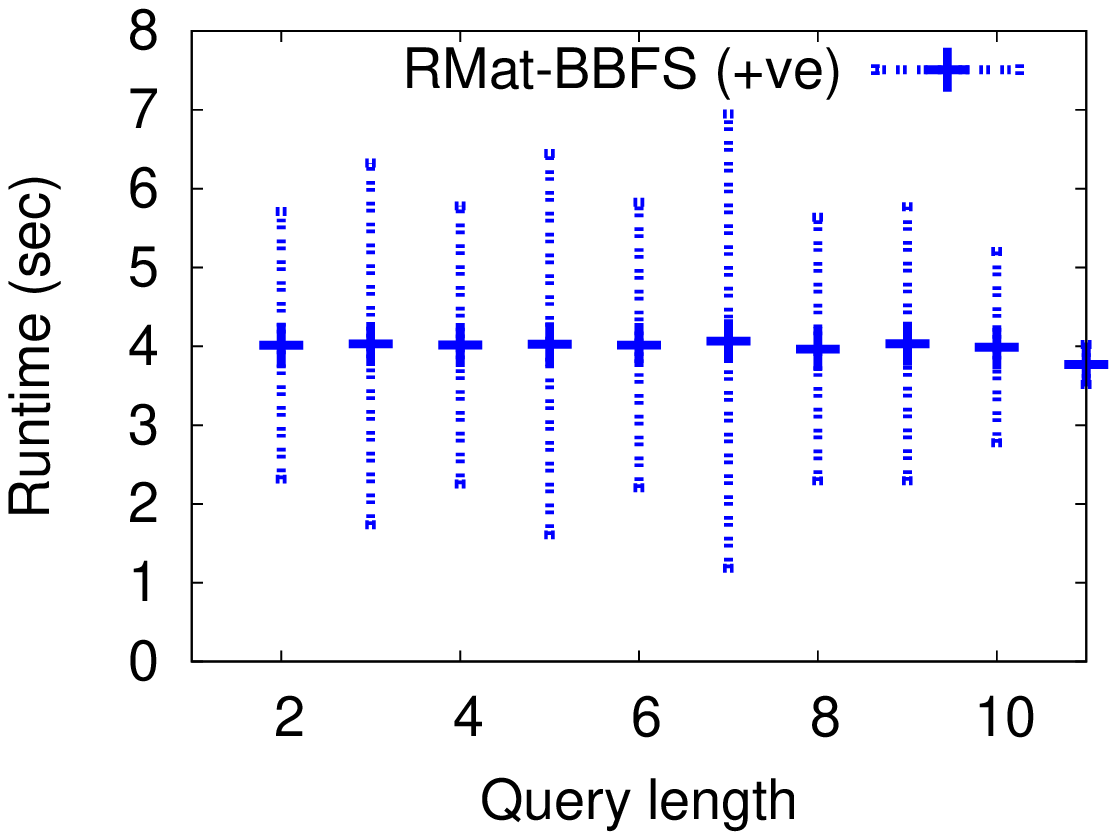}
    }
    \subfigure[]{
      \label{fig:rmat_neg_bfs}
		  \includegraphics[scale=0.43]{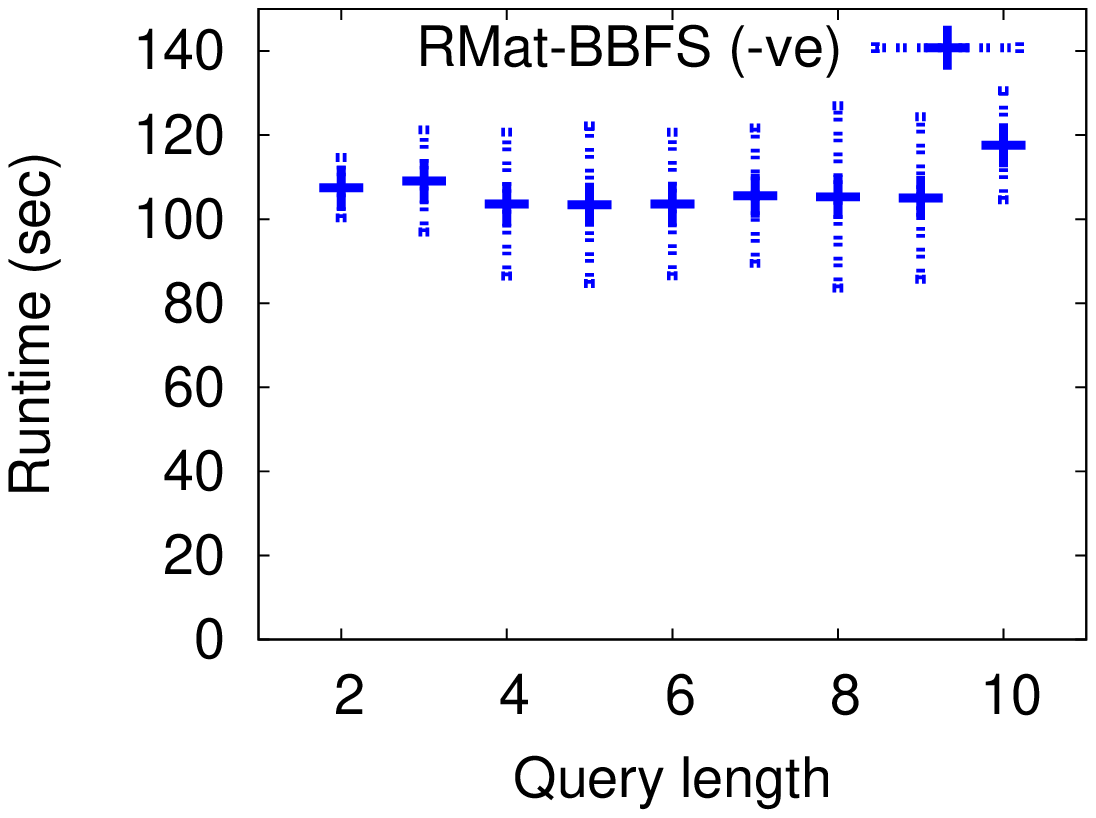}
    }
  }
  \centerline{
    \subfigure[]{
      \label{fig:rmat_fdfs}
		  \includegraphics[scale=0.43]{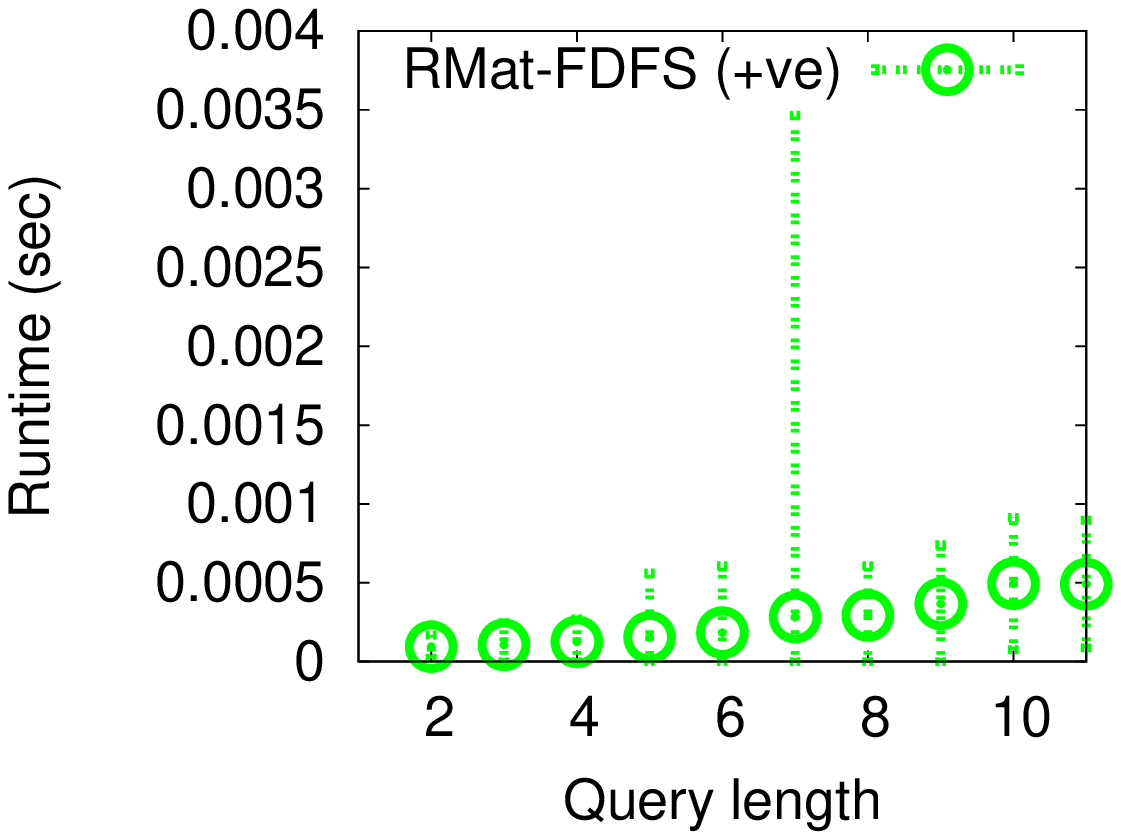}
    }
    \subfigure[]{
      \label{fig:rmat_neg_fdfs}
		  \includegraphics[scale=0.43]{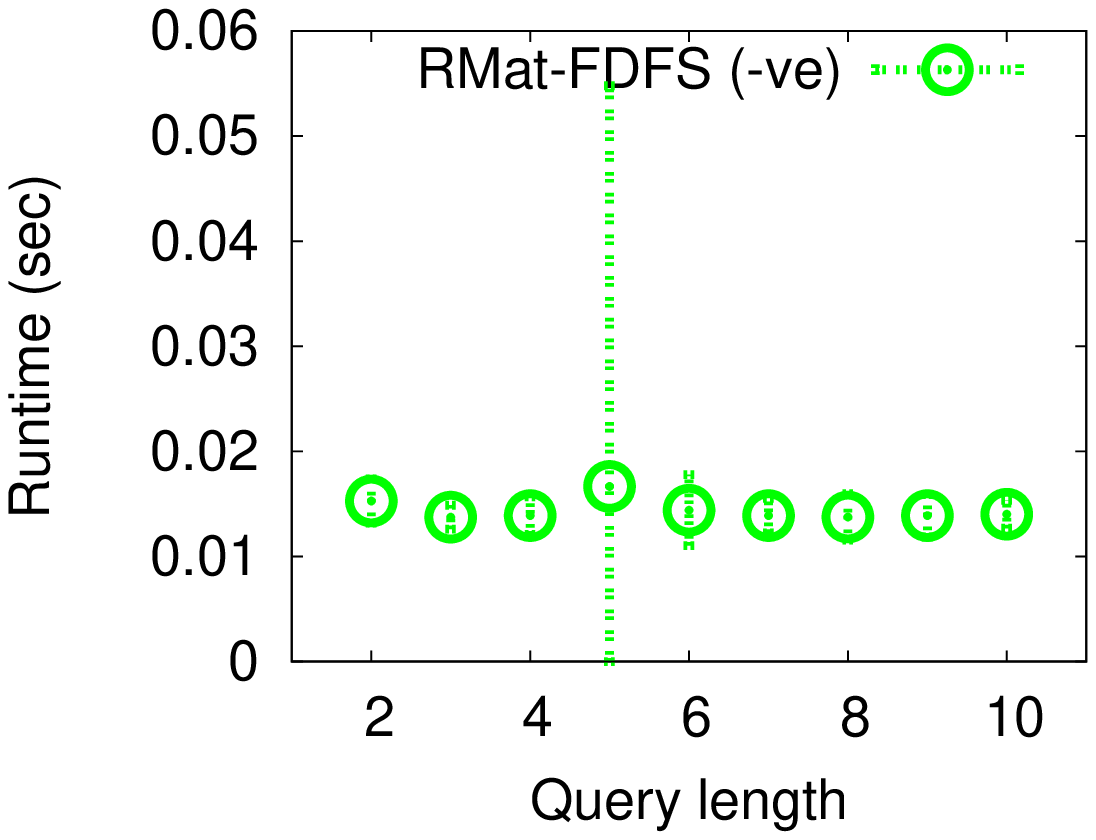}
    }
  }
\centerline{
\subfigure[]{
      \label{fig:rmat_dfs}
	  \includegraphics[scale=0.43]{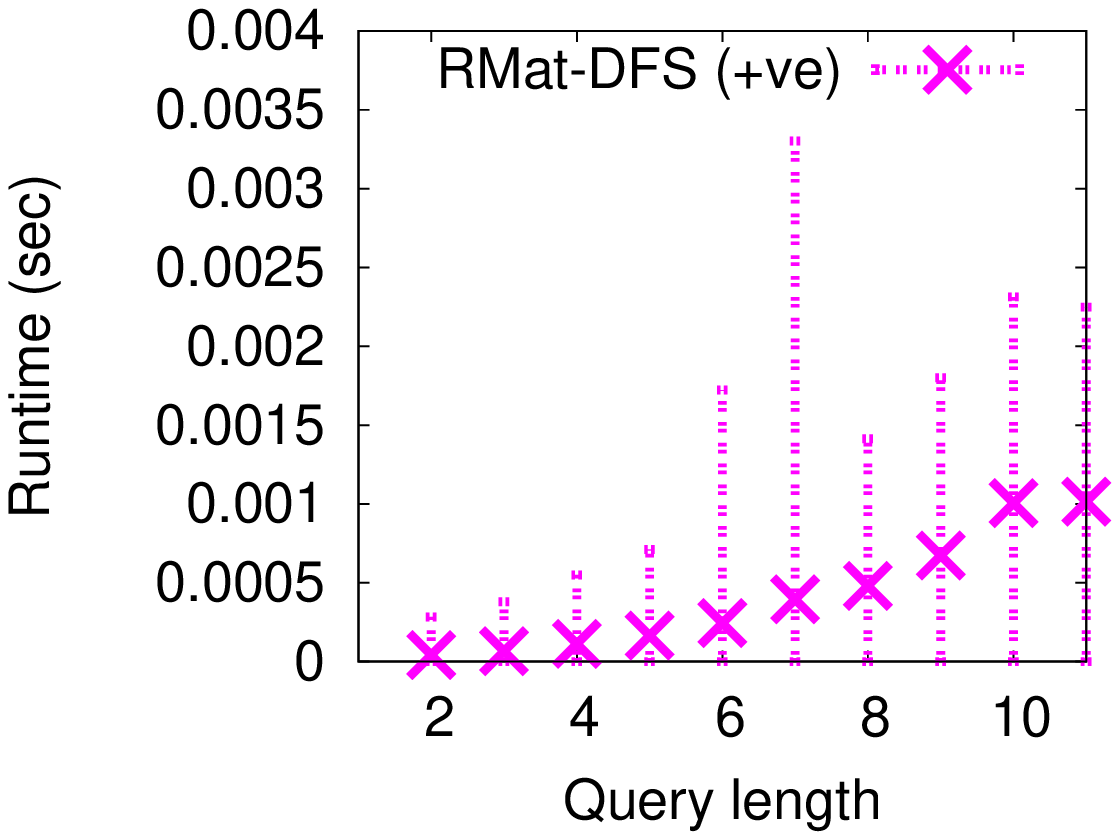}
    }
    \subfigure[]{
      \label{fig:rmat_neg_dfs}
	  \includegraphics[scale=0.43]{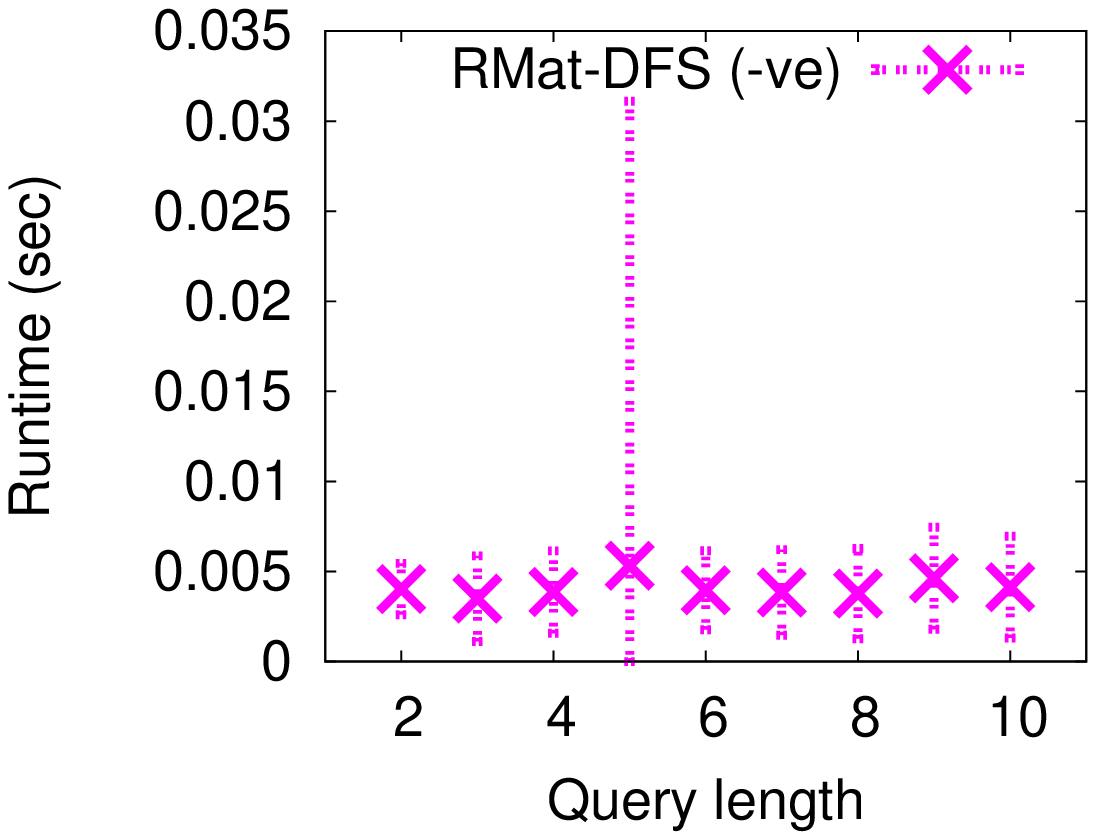}
    }
}
\caption{Mean runtime with standard deviation for varying query sizes for RMat dataset.
Row 1: BitPath, Row 2: BBFS, Row 3: FDFS, Row 4: DFS}
\label{fig:rmat_q}
\end{figure}

\begin{figure}[!ht]
  \centerline{
    \subfigure[]{
      \label{fig:uniprot_bitpath}
		  \includegraphics[scale=0.43]{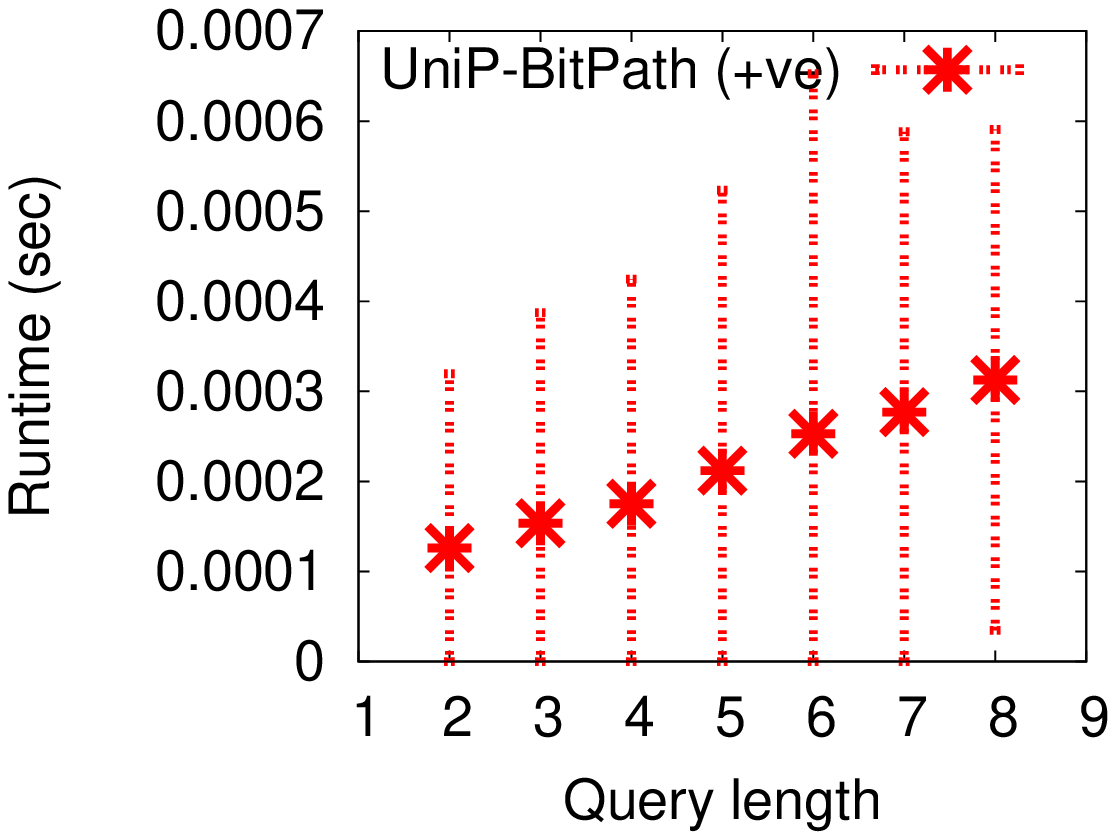}
    }
    \subfigure[]{
      \label{fig:uniprot_neg_bitpath}
	  \includegraphics[scale=0.43]{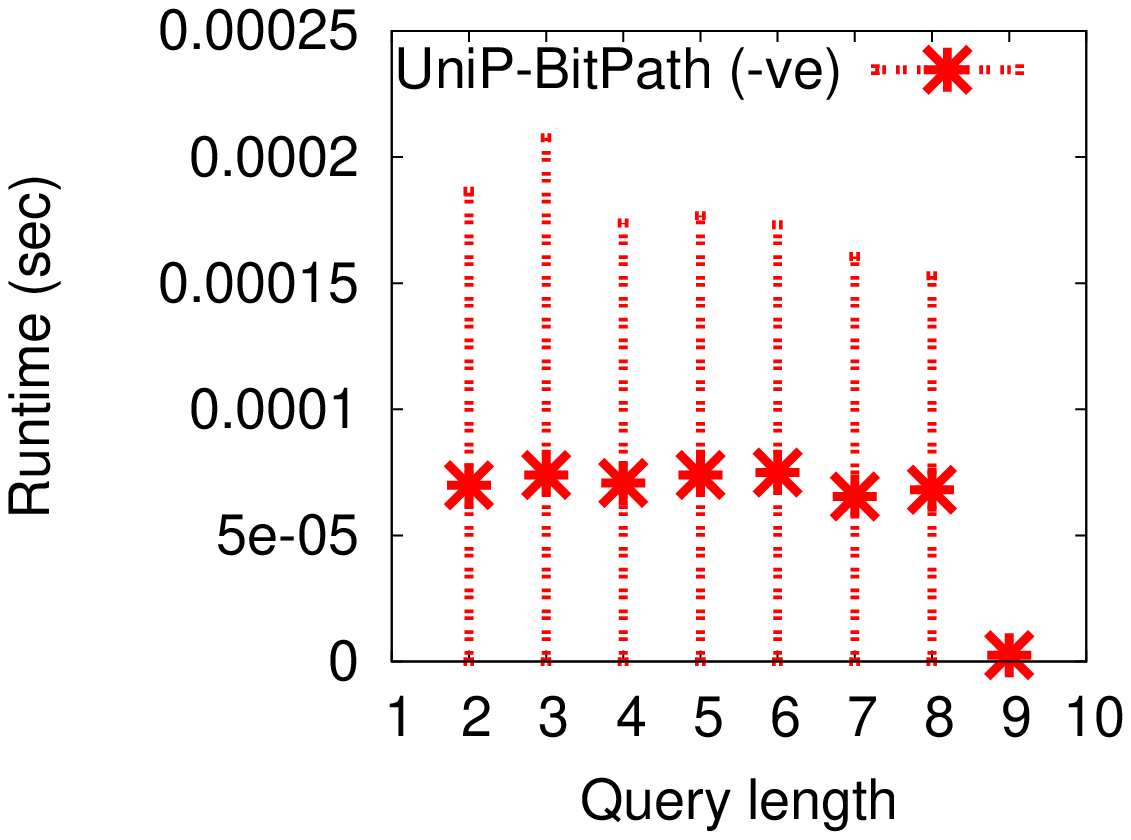}
    }
  }
  \centerline{
    \subfigure[]{
      \label{fig:uniprot_bfs}
		  \includegraphics[scale=0.43]{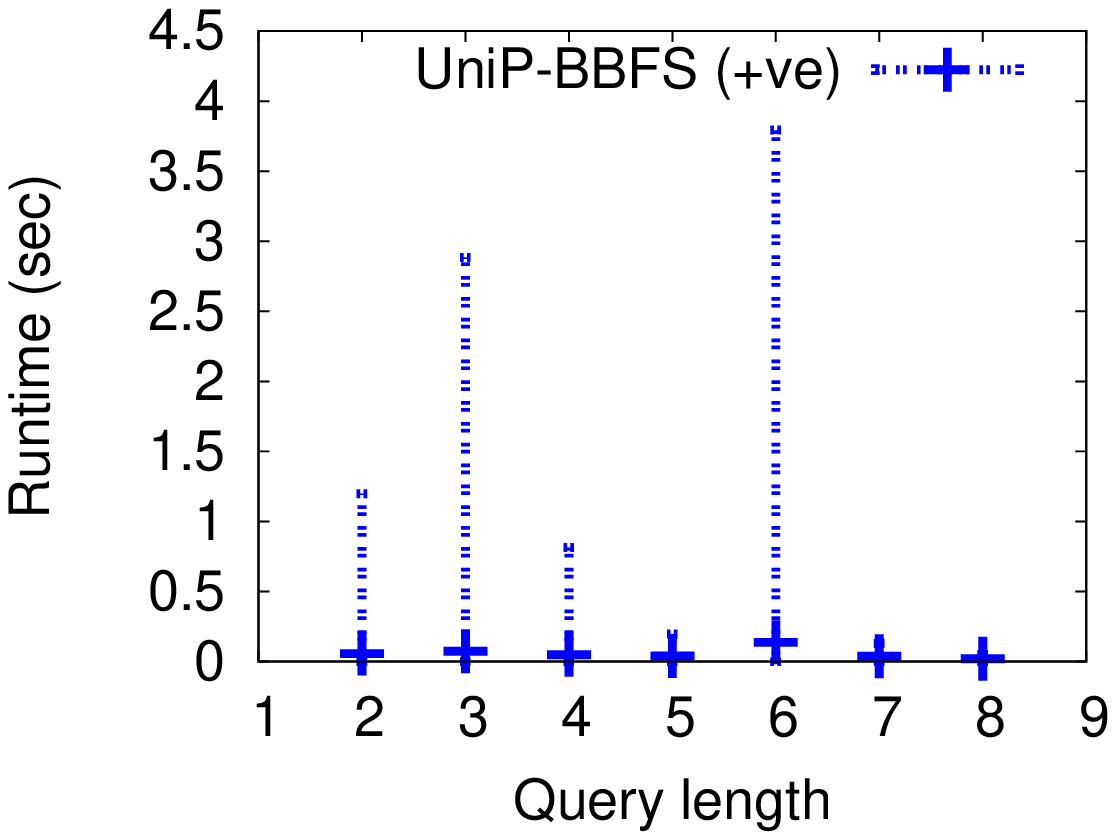}
    }
    \subfigure[]{
      \label{fig:uniprot_neg_bfs}
	  \includegraphics[scale=0.43]{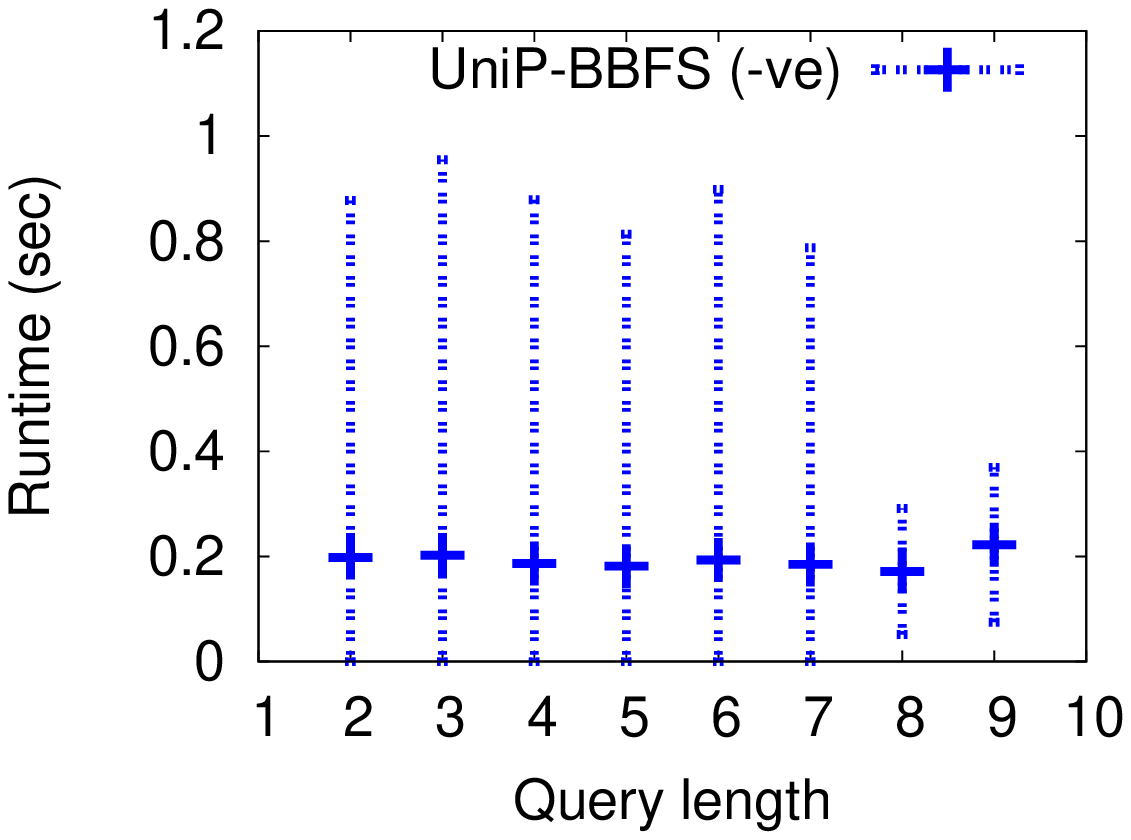}
    }
  }
  \centerline{
    \subfigure[]{
      \label{fig:uniprot_fdfs}
		  \includegraphics[scale=0.43]{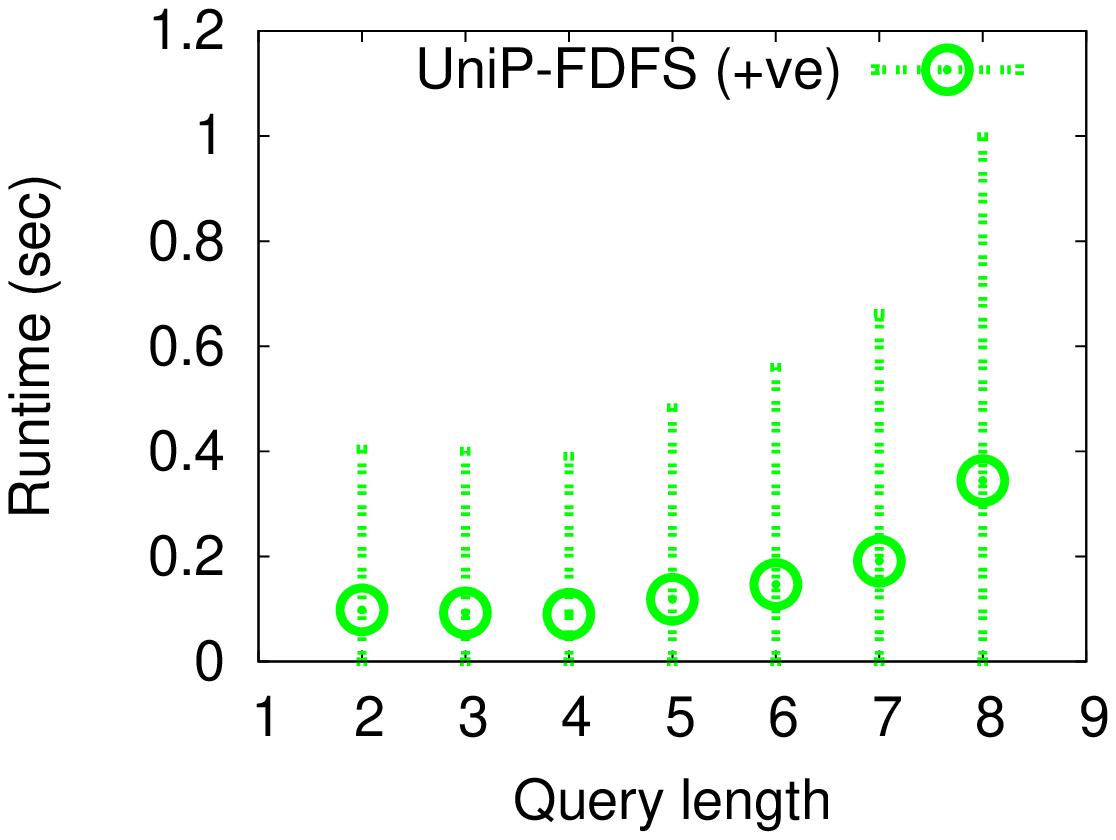}
    }
    \subfigure[]{
      \label{fig:uniprot_neg_fdfs}
	  \includegraphics[scale=0.43]{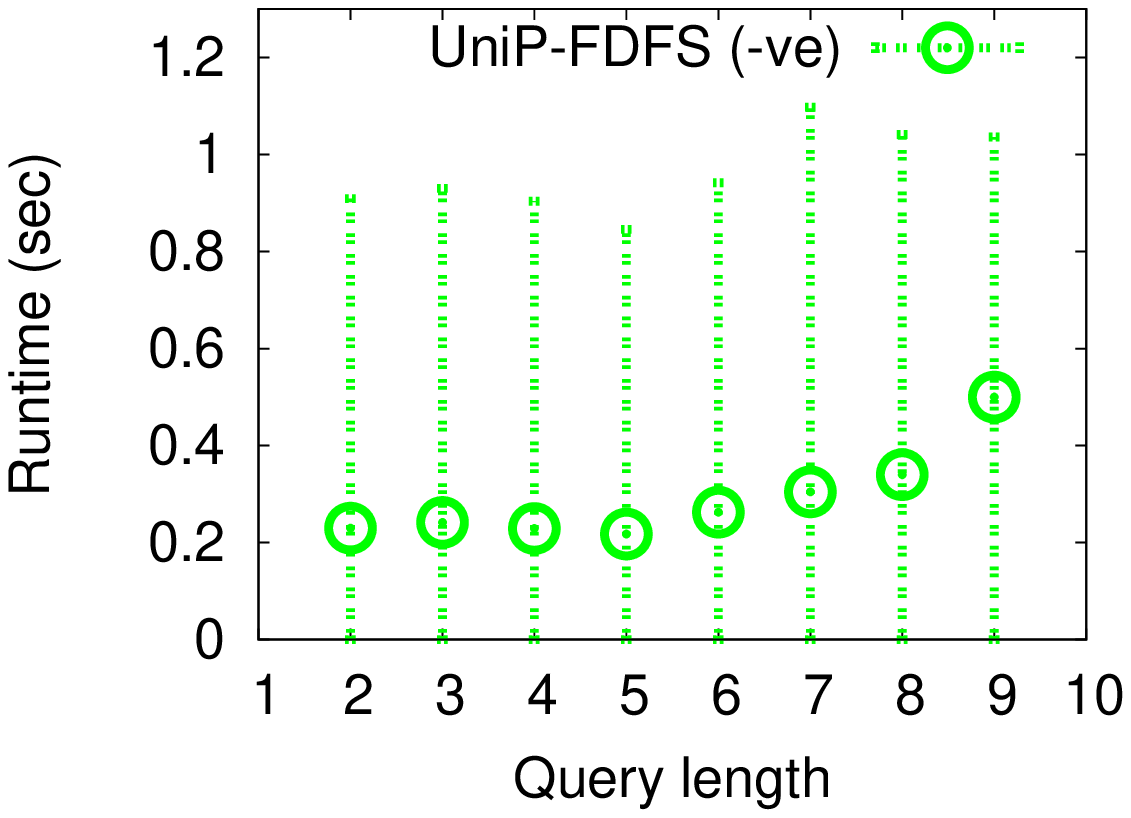}
    }
  }
\centerline{
    \subfigure[]{
      \label{fig:uniprot_dfs}
		  \includegraphics[scale=0.43]{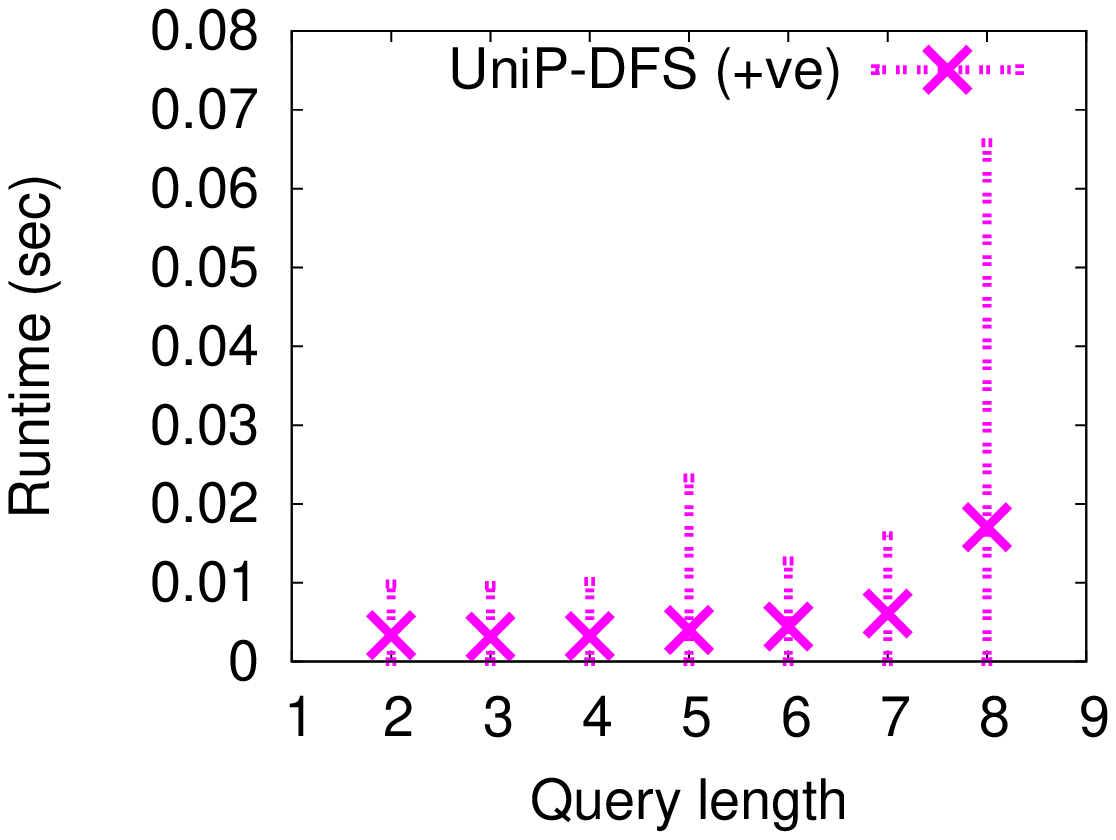}
    }
    \subfigure[]{
      \label{fig:uniprot_neg_dfs}
	  \includegraphics[scale=0.43]{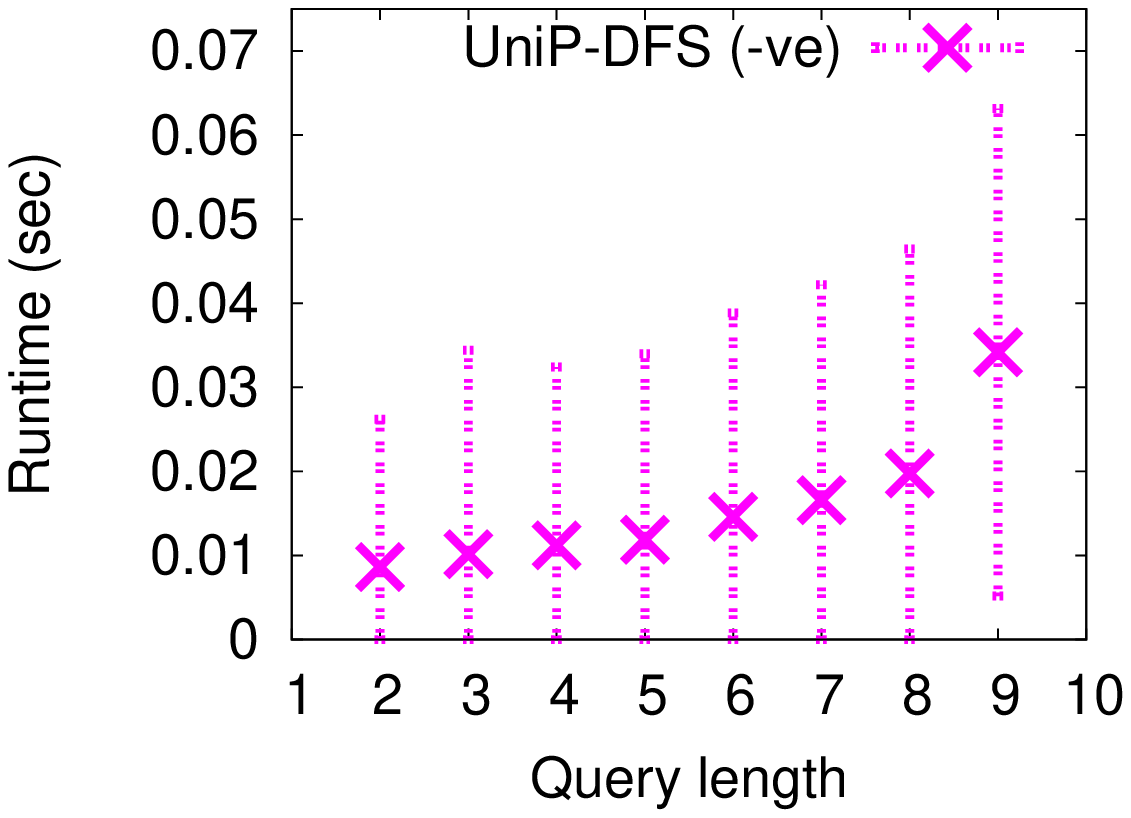}
    }
}
\caption{Mean runtime with standard deviation for varying query sizes for UniProt dataset.
Row 1: BitPath, Row 2: BBFS, Row 3: FDFS, Row 4: DFS}
\label{fig:uniprot_q}
\end{figure}

\begin{figure}[!ht]
  \centerline{
    \subfigure[]{
      \label{fig:sweto_bitpath}
	  \includegraphics[scale=0.43]{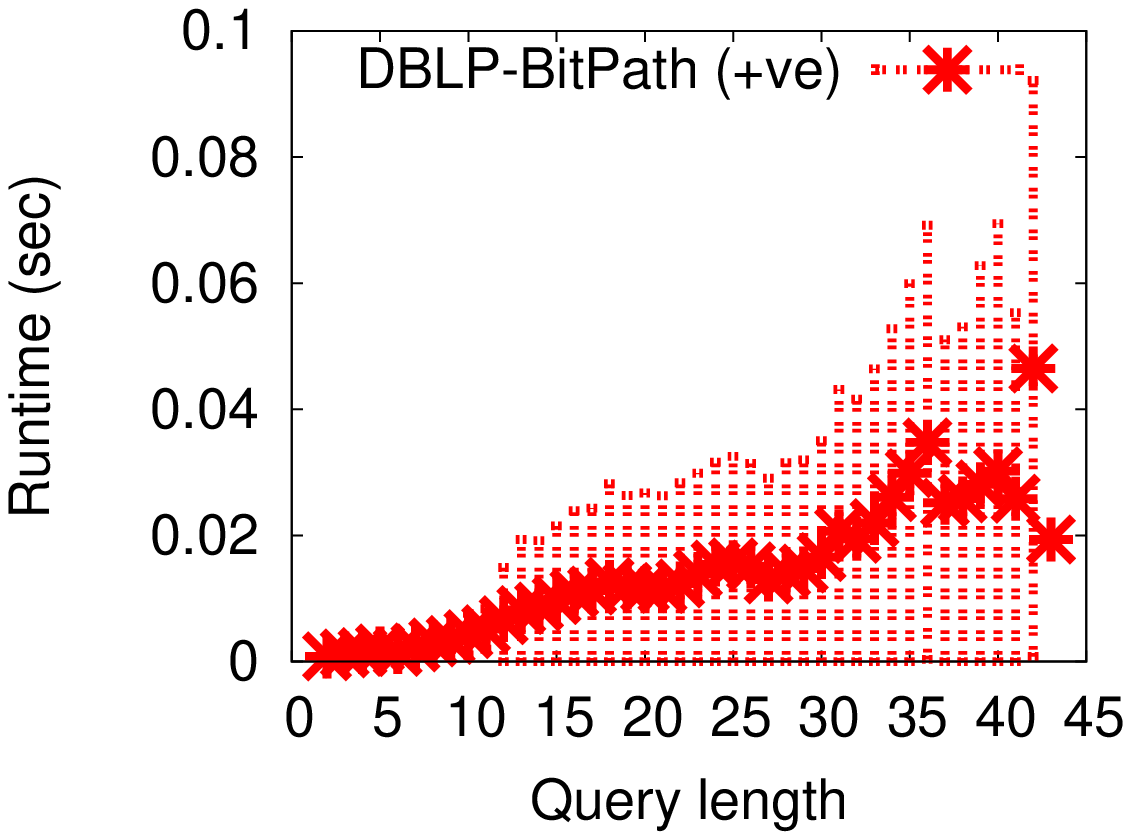}
    }
    \subfigure[]{
      \label{fig:sweto_neg_bitpath}
	  \includegraphics[scale=0.43]{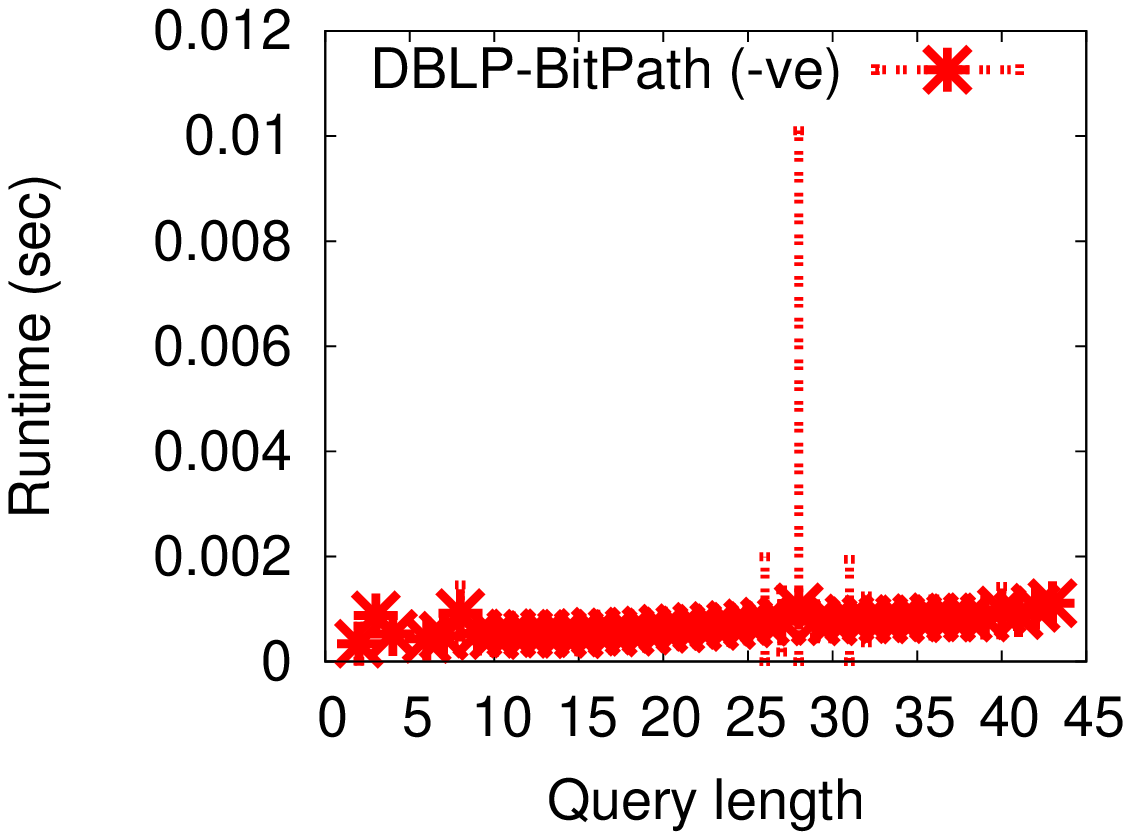}
    }
  }
\centerline{
    \subfigure[]{
      \label{fig:sweto_bfs}
	  \includegraphics[scale=0.43]{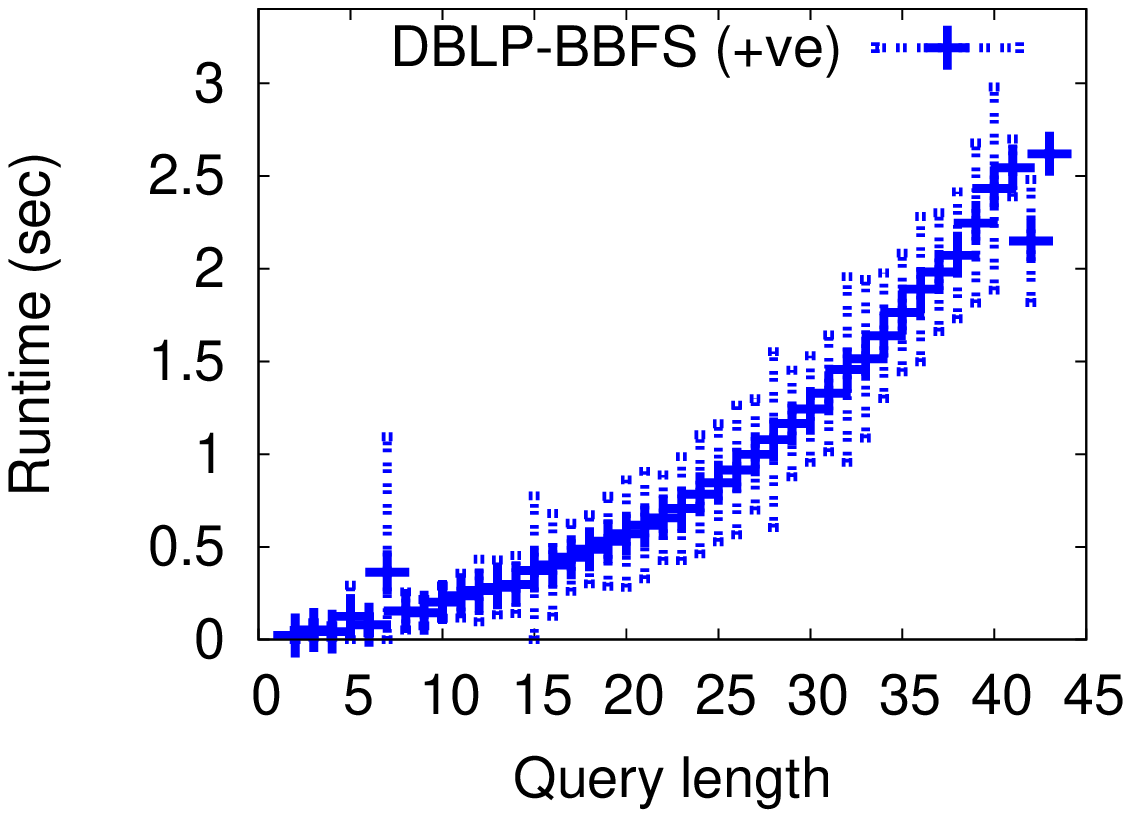}
    }
    \subfigure[]{
      \label{fig:sweto_neg_dfs}
	  \includegraphics[scale=0.43]{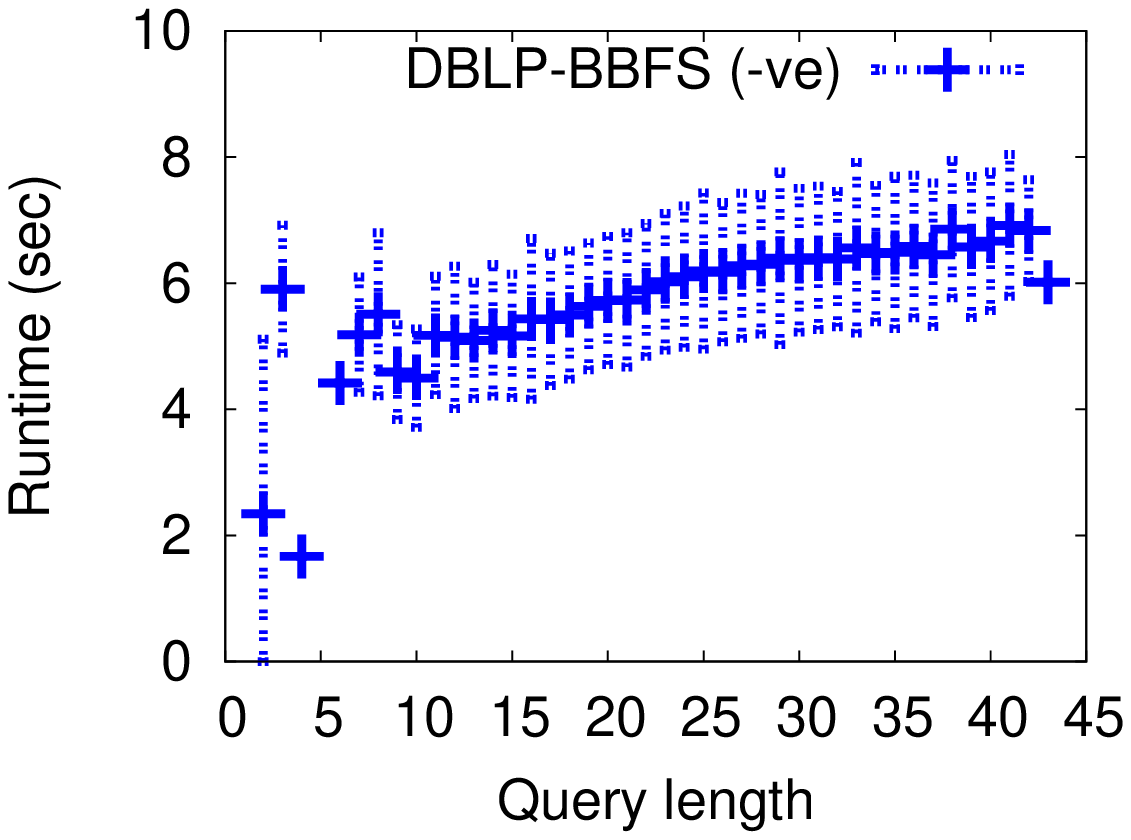}
    }
}
\caption{Mean runtime with standard deviation for varying query sizes for SwetoDBLP dataset.
Row 1: BitPath, Row 2: BBFS}
\label{fig:sweto_q}
\end{figure}

We ran the 50K positive and negative queries each on all four methods --
BitPath, DFS, F-DFS, and B-BFS. For long running queries, we set a threshold of 15
minutes/query, i.e., if a query ran for more than 15 minutes, it is abandoned.
Note the following:

\textbf{--} For SwetoDBLP's positive as well as negative queries, DFS as well as
F-DFS methods took significant time to finish on most queries. Hence we
abandoned evaluating the queries using the two DFS methods on SwetoDBLP.

\textbf{--} For SwetoDBLP's negative queries, B-BFS was taking significantly
longer time to finish (more than 24 hours for 15K queries), hence we
have shown results on only 15K queries.

\textbf{--} For R-Mat's positive queries, B-BFS was taking long time to finish,
hence after running the process for 24 hours, we abandoned further query
processing and have shown results on 43K queries.

\textbf{--} For R-Mat's negative queries, B-BFS was extremely slow taking more
than 15 minutes for most queries, hence we have shown results only on 980
queries.

Figures \ref{fig:rmat_q}, \ref{fig:uniprot_q}, and \ref{fig:sweto_q} show summarized results of Evaluation Metric (1)
as outlined in the previous subsection. Each graph shows
the average runtime (by the tick on the vertical line)
for a group of queries with same query length along with the standard deviation
for that specific group.
For example, row 1 column 1 shows BitPath's performance on R-Mat dataset for
\textit{positive} queries. It shows that for a group of queries of length 9, the
average query runtime is 0.0003 sec and the standard deviation for this group of
queries is 0.0005 sec.
Note that the scale of \textit{Y-axis} on each graph is different, e.g., the \textit{Y-axis}
for graph in row 5 column 1 for BitPath's performance over SwetoDBLP
is from \textit{0--0.1} whereas for graph in row 5 column 2 for B-BFS' performance on
SwetoDBLP is from \textit{0--3}.

\textbf{Further analysis of the results shows:}
B-BFS has inferior performance on the R-Mat graphs. This can be attributed to the
\textit{flatter} structure with not many long paths in R-Mat graphs (refer to
Table \ref{tbl:datasets} which shows that the ``largest depth'' of a node is
12), whereas SwetoDBLP graph of similar size and number of nodes has
\textit{deeper} structure with a lot of \textit{interleaved} paths (``largest depth'' of
a node in SwetoDBLP is 66).
The flatter structure of R-Mat favors DFS method
over B-BFS while on the other hand DFS method is inferior on SwetoDBLP due to its
deeper structure. Analysis of UniProt graphs shows that it contains many
\textit{disconnected} subgraphs, as a result of which B-BFS, F-DFS, as well as DFS
fair well on this graph.

B-BFS method delivers acceptable performance on SwetoDBLP graph on positive
queries.
Note that this was possible due to our \textit{optimized} version of B-BFS (ref. Section
\ref{sec:compmethods}). The naive bidirectional BFS method was not able to deliver same
performance. But as it can be noted, the performance of B-BFS method deteriorates as
the length of the query increases.
For negative queries, B-BFS method suffers on
SwetoDBLP graph as it has to explore the entire subgraph between source and
destination node. Note that on all 3 datasets with varying characteristics,
BitPath delivers uniform performance on positive as well as negative queries.
On average BitPath's performance is 50 to 1000 times better for positive
queries, and 1000 to 100k times better for negative queries compared to the baseline
methods.

\subsection{BitPath Index Size and Construction Time} \label{sec:indxsize}

The BitPath index construction time for R-Mat, UniProt, and SwetoDBLP datasets
is 933 sec, 292 sec, and 809 sec respectively. Since the procedure of merging
strongly connected components (SCCs) is same across all methods, the index
construction time here does not include identification and merging of SCCs.

The cumulative on-disk size of N-SUCC-E, N-PRED-E, EL-ID and EID indexes for R-Mat,
UniProt, and SwetoDBLP datasets are 3.7 GB, 5 GB, and 3 GB, whereas
the on-disk size of these graphs is  243 MB, 394 MB, and 232 MB respectively.
The uncompressed size of N-SUCC-E and N-PRED-E bit-vector indexes
for R-Mat, UniProt, and SwetoDBLP graphs would have been 15,269 GB, 37,466 GB and 18,255 GB respectively
(since each node has a bit-vector index of successor and predecessor edges, uncompressed size of these indexes in bytes
would be \#nodes $\times$ \#edges $\times$ 2 / 8).
Also, if N-SUCC-E and N-PRED-E indexes were stored as pure integer arrays
instead of compressed bit-vectors, they would have taken 2130 GB, 738 GB, and 
20 GB for R-Mat, UniProt, and SwetoDBLP graphs respectively.
Note that this size is \textit{excluding} the size of EL-ID and EID indexes, whereas the cumulative
BitPath index sizes given above include size of all 4 indexes (N-SUCC-E, N-PRED-E, EL-ID, EID).

Thus we have shown that our approach of numbering the edges and applying run-length-encoding
on the N-SUCC-E and N-PRED-E indexes (ref. Section \ref{sec:index}) reduces the on-disk size of indexes by a factor of
7--1000 over naive indexing and storage methods.

%% file: conclusion.tex
\section{Conclusion}
In this paper we have addressed \textit{label-order
constrained} reachability queries. This problem is of specific interest for large graphs with diverse
relationships (i.e., large number of edge labels in the graph).
Path indexing methods for the constrained reachability or regular path queries
work well on smaller graphs, but they often face scalability issues for
large real life graphs. Similarly, the complexity of indexing all paths prohibits
its use in practice for graphs which do not assume tree-structure.

We propose a method of building \textit{light-weight} indexes on
graphs using {\it compressed bit-vectors}. As shown in our evaluation, compressed
bit-vectors reduce the total size of the indexes significantly.
Our \textit{divide-and-conquer}
algorithm along with \textit{greedy-pruning} strategy delivers more uniform
performance across graphs of different structural characteristics. We have
evaluated our method over graphs of more than 6 million
nodes and 22 million edges -- to the best of our knowledge, the largest among the published
literature in the context of path queries on graphs.
In the future, we plan to improve this method to process a wider range of path
expressions and also to incorporate ways of enumerating actual path description.